\documentclass[paper]{JHEP3}

\usepackage{amsmath,amssymb}
\usepackage[dvips]{graphics}
\usepackage{epsfig}
\usepackage{psfrag,comment}

\newcommand{\be}{\begin{equation}}
\newcommand{\ee}{\end{equation}}
\newcommand{\bea}{\begin{eqnarray}}
\newcommand{\eea}{\end{eqnarray}}
\newcommand{\bean}{\begin{eqnarray*}}
\newcommand{\eean}{\end{eqnarray*}}
\newcommand{\ba}{\begin{aligned}}
\newcommand{\ea}{\end{aligned}}

\newcommand{\sectiono}[1]{\section{#1}\setcounter{equation}{0}}

\newcommand{\CC}{{\cal C}}

\newcommand{\CK}{{\cal K}}

\newcommand{\CO}{{\cal O}}

\def\BZ{{\mathbb Z}}

\def\BC{{\mathbb C}}

\def\r{\right\rangle}

\def\1{\mathbf{1}}
\def\0{|\1\r}

\newcommand{\p}{\partial}
\newcommand{\tr}{{\rm Tr}}
\newcommand{\rme}{{\rm e}}
\newcommand{\rmi}{{\rm i}}
\newcommand{\rmd}{{\rm d}}
\newcommand{\re}{{\rm e}}
\newcommand{\ri}{{\rm i}}
\newcommand{\rd}{{\rm d}}

\newcommand{\figref}[1]{Fig.~\protect\ref{#1}}

\newdimen\tableauside\tableauside=1.0ex
\newdimen\tableaurule\tableaurule=0.4pt
\newdimen\tableaustep
\def\phantomhrule#1{\hbox{\vbox to0pt{\hrule height\tableaurule width#1\vss}}}
\def\phantomvrule#1{\vbox{\hbox to0pt{\vrule width\tableaurule height#1\hss}}}
\def\sqr{\vbox{%
  \phantomhrule\tableaustep
  \hbox{\phantomvrule\tableaustep\kern\tableaustep\phantomvrule\tableaustep}%
  \hbox{\vbox{\phantomhrule\tableauside}\kern-\tableaurule}}}
\def\squares#1{\hbox{\count0=#1\noindent\loop\sqr
  \advance\count0 by-1 \ifnum\count0>0\repeat}}
\def\tableau#1{\vcenter{\offinterlineskip
  \tableaustep=\tableauside\advance\tableaustep by-\tableaurule
  \kern\normallineskip\hbox
    {\kern\normallineskip\vbox
      {\gettableau#1 0 }%
     \kern\normallineskip\kern\tableaurule}%
  \kern\normallineskip\kern\tableaurule}}
\def\gettableau#1{\ifnum#1=0\let\next=\null\else
\squares{#1}\let\next=\gettableau\fi\next}
\preprint{
{\small{\textsf{CERN-PH-TH/2008-186}}}}

\title{Multi--Instantons and Multi--Cuts}

\author{Marcos Mari\~no$^1$, Ricardo Schiappa$^2$ and Marlene Weiss$^{3,4}$
\\
$^1$Section de Math\'ematiques et D\'epartement de Physique Th\'eorique,\\
Universit\'e de Gen\`eve, CH--1211 Gen\`eve, Switzerland\\
\\
$^2$CAMGSD, Departamento de Matem\'atica, \\
Instituto Superior T\'ecnico, 1049--001 Lisboa, Portugal\\
\\
$^3$Theory Division, Department of Physics,\\ 
CERN, CH--1211 Gen\`eve 23, Switzerland\\
\\
$^4$ITP, ETH Z\"urich,\\
CH--8093 Z\"urich, Switzerland\\
\\
\email{marcos.marino@unige.ch}, \quad
\email{schiappa@math.ist.utl.pt}, \quad
\email{marlene.weiss@cern.ch}
}
\abstract{
We discuss various aspects of multi--instanton configurations in generic multi--cut matrix models. Explicit formulae are presented in the two--cut case and, in particular, we obtain general formulae for multi--instanton amplitudes in the one--cut matrix model case as a degeneration of the two--cut case. These formulae show that the instanton gas is ultra--dilute, due to the repulsion among the matrix model eigenvalues. We exemplify and test our general results in the cubic matrix model, where multi--instanton amplitudes can be also computed with orthogonal polynomials. As an application, we derive general expressions for multi--instanton contributions in two--dimensional quantum gravity, verifying them by computing the instanton corrections to the string equation. The resulting amplitudes can be interpreted as regularized partition functions for multiple ZZ--branes, which take into full account their back--reaction on the target geometry. Finally, we also derive structural properties of the trans--series solution to the Painlev\'e I equation.
}


\begin{document}



\vfill

\eject

\sectiono{Introduction}

The computation of instanton effects in generic matrix models is an important problem which has many applications. For example, matrix models describe, in the so--called double--scaling limit, noncritical or minimal (super)string theories, and nonperturbative effects in the matrix model are related to the nonperturbative structure of these string theories (see \cite{dfgz} for an excellent review). Indeed, these instanton effects have been interpreted in terms of D--brane configurations (see for example \cite{martinec, akk}) and were also instrumental for the discovery of D--branes in critical string theory \cite{boundaries}. Such connections motivated the first computations of nonperturbative effects in matrix models, in the pioneering works of David and of Shenker \cite{david1, david2, davidvacua, shenker}, and recent progress in understanding the continuum side \cite{zz} has further stimulated the analysis of instanton effects also on the matrix model side (such as, for example, \cite{lvm,st}).

However, most of the progress in the study of these effects in matrix models has focused on the double--scaling limit, \textit{i.e.}, when the matrix model is in the vicinity of a critical point. The study of nonperturbative effects for generic values of the coupling constants is also very important \textit{per se}, and recently it has been further stimulated by the discovery that matrix models, away from criticality, describe in some cases topological string theory on certain non--compact Calabi--Yau threefolds \cite{dv, mm, remodeling}. With this motivation in mind, two different approaches have been very recently developed with the goal of computing instanton effects in generic matrix models off--criticality. The first approach, presented in \cite{msw}, builds on previous computations in the double--scaling limit \cite{david2,lvm,iy}, extending them away from criticality. It is based on saddle--point techniques, and essentially computes the instanton amplitudes directly in terms of a saddle--point integral. The result can be expressed in terms of generating functions of correlation functions in the perturbative theory. Results for the one--instanton amplitude at two loops, in matrix models with a single cut, were obtained with this technique in \cite{msw}, but going to higher order in instanton number and/or the number of loops seems rather complicated with this technique. The second approach, presented in \cite{mmnp}, is based on the method of orthogonal polynomials, and generalizes the techniques first introduced in \cite{biz} in order to deal with instanton calculus. A clear advantage of this approach with respect to the latter is that it allows for a clean evaluation of multi--instanton amplitudes. However, since not all phases of matrix models can be appropriately described with orthogonal polynomials (the multi--cut case being a notorious example), the methods of \cite{mmnp} can not be universally applied, although they are computationally very powerful. As such, both \cite{msw} and \cite{mmnp} yield results only for matrix models with a single cut.

In this paper we address multi--instanton configurations in multi--cut matrix models, hoping to go one step further with respect to \cite{msw, mmnp}. It turns out that, for a generic multi--cut matrix model background with \textit{fixed} filling fractions (\textit{i.e.}, with a fixed number of eigenvalues in the cuts), a multi--instanton configuration is just any \textit{other} choice of filling fractions. Formal expressions for the instanton amplitudes can then be obtained just by expanding the partition function around a generic filling fraction, making in this way contact with the results of \cite{bde, eynard}. Therefore, the problem of obtaining multi--instanton amplitudes in the generic multi--cut matrix model is solved in principle by the problem of computing the partition function for a generic multi--cut background, and the complexity of evaluating nonperturbative effects reduces to the complexity of computing such a generic partition function.

Herein, we shall consider generic multi--cut backgrounds where the eigenvalues can sit both at local minima as well as local maxima, which are the relevant configurations when addressing topological strings. Such a generic, unstable vacuum has tachyonic directions, therefore nonperturbative contributions to the perturbative partition function can be exponentially suppressed or exponentially enhanced. However, we shall see that it is still possible to make sense of the instanton expansion of the free energy by using the properties of the saddle--point $1/N$ expansion of the matrix model, and in particular of the theta--function representation discussed in \cite{bde, eynard}.

In our approach, multi--instanton amplitudes in the one--cut model can be obtained as particular degeneration limits of the multi--cut case, by expanding around a background in which all but one cuts are empty. The resulting amplitude is, however, singular and it has to be appropriately regularized. We propose a natural regularization which leads to a general formula for multi--instanton amplitudes in one--cut matrix models. We then perform various tests of this formula in two examples: the cubic matrix model and its double--scaling limit, namely, two--dimensional quantum gravity. In both cases multi--instanton effects can be independently computed by using trans--series expansions of the recursion relations defining the free energy. In 2d gravity, this recursion takes the form of the famous Painlev\'e I equation for the specific heat. The results agree with our general formula, therefore justifying our regularization. Notice that the appearance of divergences in multi--instanton amplitudes has been noticed before, in the particular case of the double--scaling limit, where they correspond to singularities in the amplitudes for multiple, coincident ZZ--branes \cite{annulusZZ, st}. However, it is also known that these divergences should be absent once the back--reaction of the ZZ--branes is appropriately taken into account. Our regularization procedure can then be interpreted as the \textit{correct} prescription to incorporate this back--reaction and, in particular, our results concerning multi--instantons in 2d gravity are regularized amplitudes for multiple ZZ--branes in the $(2,3)$ model coupled to gravity.

The organization of this paper is as follows. We begin in section 2 by reviewing both the general structure of multi--cut matrix models and the definition of the multi--instanton configurations we shall address. The general partition function of the matrix model is defined as a sum over all possible filling fractions and this leads to the theta--function representation of the partition function, which we relate to the multi--instanton story. In the following section 3 we then review the solution of multi--cut matrix models, in the large $N$ limit, with a special emphasis on presenting explicit formulae for the two--cut case. In particular, we compute, in the two--cut case, all quantities that we earlier addressed in section 2. In this way, we are ready to shift gears in section 4 and consider multi--instantons in the one--cut matrix model as a particular, degenerate case of our previous two--cut considerations. We compute multi--instanton amplitudes up to two loops and discuss the nature of the instanton gas in the matrix model. We then address some examples, such as the cubic matrix model and 2d quantum gravity. Both cases may also be tackled by making use of trans--series techniques, which provide an extra check on our results. We end with a concluding section, and also collect some more technical details in a few appendices.

\sectiono{Multi--Instantons and Multi--Cut Matrix Models: General Aspects}

In this section we discuss general aspects of multi--cut matrix models and multi--instantons. We first present a quick review of multi--cut matrix models, following \cite{davidvacua, bde, akemann, leshouches} (more technical results will be discussed in the next section). Then, we relate multi--cut matrix models to multi--instanton configurations in a general setting.

\subsection{Multi--Cut Matrix Models}

We recall that the one--matrix model partition function is
\be\label{pf}
Z_N = \frac{1}{{\mathrm{vol}} \left[ U(N) \right]} \int \rmd M\, \exp \left( - \frac{1}{g_s} \tr\, V(M) \right),
\ee
\noindent
where $V(x)$ is a potential which we shall take to be a polynomial. This can be written in diagonal gauge in terms of eigenvalues as
\be
Z_N = \frac{1}{N!} \int \prod_{i=1}^N  \frac{\rmd \lambda_i}{2\pi} \, \Delta^2(\lambda)\, \exp \left( - \frac{1}{g_s} \sum_{i=1}^N V(\lambda_i) \right),
\ee
\noindent
where $\Delta(\lambda)$ is the Vandermonde determinant. As usual, we are interested in studying the model at large $N$ but keeping the 't~Hooft coupling 
\be\label{thooft}
t=Ng_s
\ee
\noindent
fixed. In this limit the model is described by the density of eigenvalues
\be\label{density}
\rho(\lambda) = {1\over N}\, \Big\langle \tr \, \delta(\lambda-M) \Big\rangle.
\ee

\FIGURE[ht]{
\leavevmode
\centering
\epsfysize=4.5cm
\hspace{3cm}
\epsfbox{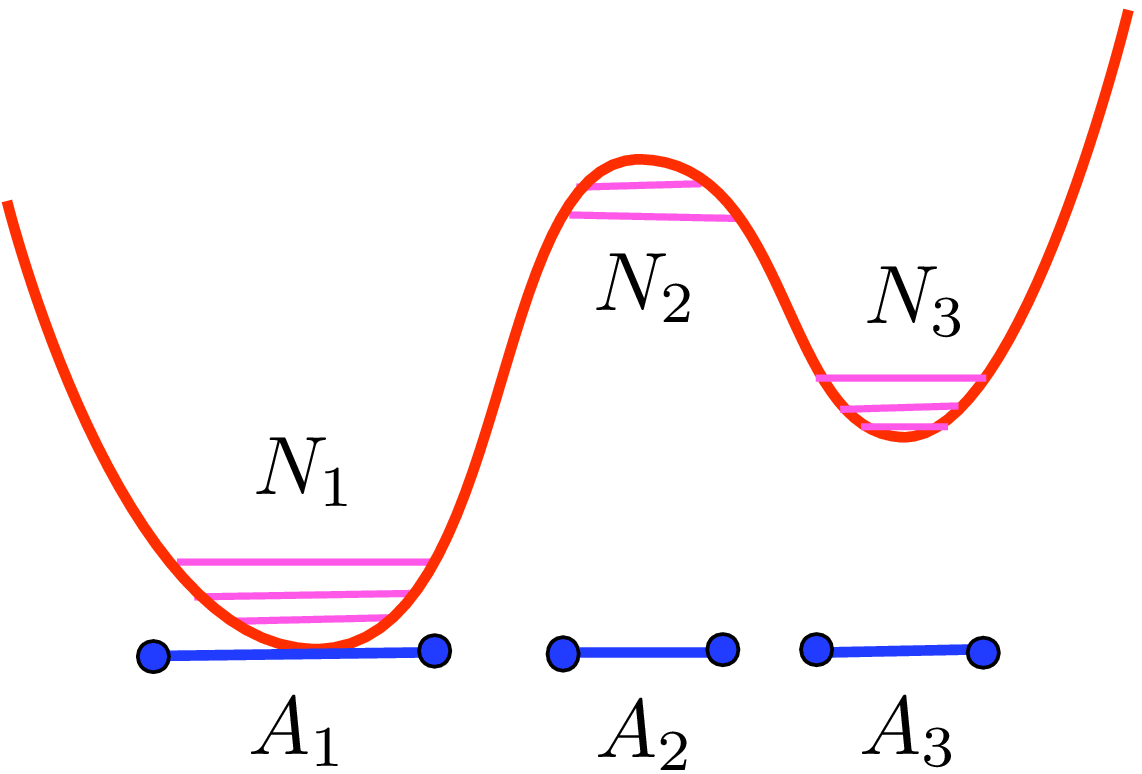}
\hspace{3cm}
\caption{An example of a potential leading to a three--cut solution. The $N$ eigenvalues split into three sets, $N_I$, $I=1,2,3$, and they sit around the extrema of the potential. The support of the density of eigenvalues is the union of the three intervals $A_I$, $I=1,2,3$.}
\label{multicut}
}

Let us assume that the potential $V(x)$ has $s$ extrema. The most general saddle--point solution of the model, at large $N$, will be characterized by a density of eigenvalues supported on a disjoint union of $s$ intervals 
\be\label{cint}
\CC = \bigcup_{I=1}^s A_I,
\ee
\noindent
where $A_I = [ x_{2I-1}, x_{2I} ]$ are the $s$ cuts and $x_1 < x_2 < \cdots < x_{2s}$. If $s>1$ we call this saddle--point a \textit{multi--cut solution} of the Hermitian matrix model. A potential leading to a three--cut solution is depicted in \figref{multicut}. This multi--cut saddle--point can be described in terms of an integration over eigenvalues as follows. In the $s$--cut configuration described above the $N$ eigenvalues split into $s$ sets of $N_I$ eigenvalues, $I=1, \ldots, s$. Let us denote each of these $s$ sets by
\be
\{ \lambda^{(I)}_{k_I} \}_{k_I=1, \ldots, N_I}, \qquad I=1, \ldots, s. 
\ee
\noindent
The eigenvalues in the $I$--th set sit in the interval $A_I$ around the $I$--th extremum. Along this interval, the effective potential
\be
V_{\rm eff}(\lambda) = V(\lambda) - t \int \rd \lambda' \rho(\lambda') \log|\lambda -\lambda'| 
\ee
\noindent
is constant. It is possible to choose $s$ integration contours $\CC_I$ in the complex plane, $I=1, \ldots, s$, going off to infinity along directions where the integrand decays exponentially, and in such a way that each of them passes through exactly one of the $s$ critical points (see for example \cite{fr}). The resulting matrix integral is then convergent and can be written as
\be\label{genz}
Z(N_1, \ldots, N_s) = {1 \over N_1! \cdots N_s!} \int_{\lambda^{(1)}_{k_1} \in \CC_1} \cdots \int_{\lambda^{(s)}_{k_s} \in \CC_s} \prod_{i=1}^N {\rd\lambda_i \over 2 \pi}\, \Delta^2(\lambda)\, \re^{-{1\over g_s} \sum_{i=1}^N V(\lambda_i)}.
\ee
\noindent
The overall combinatorial factor in this expression, as compared to the one in (\ref{pf}), 
is due to the fact that there are 
\be
{N! \over N_1! \cdots N_s!}
\ee
\noindent
ways to choose the $s$ sets of $N_I$ eigenvalues. Of course, when the integrand is written out in detail, it splits into $s$ sets of eigenvalues which interact among them through the Vandermonde determinant (see for example \cite{kmt}). In this paper we will assume that the filling fractions, 
\be\label{filfil}
\epsilon_I \equiv \frac{N_I}{N}= \int_{A_I} \rmd\lambda\, \rho(\lambda), \qquad I = 1,2,\ldots,s,
\ee
\noindent
are fixed, and they can be regarded as parameters, or moduli, of the model. One may also use as moduli the partial 't~Hooft couplings $t_I = t \epsilon_I = g_s N_I$, which can be written as
\be\label{thooftmoduli}
t_I = \frac{1}{4\pi\rmi} \oint_{A^I} \rmd z\, y(z),
\ee
\noindent
with $\sum_{I=1}^s t_I = t$. Here $y(z)$ is the spectral curve of the matrix model which, in the multi--cut case, describes a hyperelliptic geometry, and relates to the effective potential as $V_{\rm eff}'(z) = y(z)$. We shall be more concrete about it in the next section. Notice that, in general, the saddle--points we consider will have unstable directions. This is in fact the generic situation one finds in the applications of matrix models to topological string theory \cite{dv, mm, remodeling}. Indeed, as we shall learn in this paper, precisely the fact that we consider general saddle--points will make it possible to extract multi--instanton amplitudes from multi--cut matrix models.

The free energy of the multi--cut matrix model has a genus expansion of the form 
\be
F=\log Z=\sum_{g=0}^{\infty} F_g(t_I)\, g_s^{2g-2}. 
\ee
\noindent
In particular, the planar free energy $F_0(t_I)$ can be computed from the spectral curve by using the special geometry relation 
\be\label{bmodel}
{\partial F_0(t) \over \partial t_I} = \oint_{B_{\Lambda}^I} \rd \lambda\, y(\lambda), \qquad I=1, \ldots, s, 
\ee
\noindent
where $B_{\Lambda}^I$ is a path which goes from the endpoint of the $A_I$ cycle to the cut--off point $\Lambda$. This point may then be taken to infinity after one removes any divergent pieces from the integral. The higher $F_g (t_I)$ can also be computed explicitly, and there is a systematic, recursive procedure to obtain them, put forward in \cite{ce, eo}.

\subsection{Multi--Instantons in Matrix Models}

If one regards (\ref{genz}) as the matrix integral in a specific topological sector, characterized by the fillings $N_1, \ldots, N_s$, it is then natural to consider the general partition function as a sum over all possible arrangements of eigenvalues across the several cuts \cite{david1, davidvacua, bde, eynard}
\be\label{sumz}
Z (\zeta_1, \ldots, \zeta_s) = \sum_{N_1 +\cdots +N_s=N} \zeta_1^{N_1} \cdots \zeta_s^{N_s} Z(N_1, \ldots, N_s).
\ee
\noindent
The coefficients $\zeta_k$ may be regarded as $\theta$--parameters which lead to different $\theta$--vacua, and we refer the reader to \cite{davidvacua} for such an interpretation. The sum (\ref{sumz}) may also be regarded as a matrix integral where the $N$ eigenvalues are integrated along the contour
\be
\CC =\sum_{k=1}^s \zeta_k \CC_k\, ;
\ee
\noindent
in this case the $\theta$--parameters yield the relative weight of the different contours $\CC_k$, see \cite{davidvacua,eynard}.

A very important point is the fact that the different sectors appearing in (\ref{sumz}) can be regarded as instanton sectors of the multi--cut matrix model. Indeed, let $Z(N_1, \ldots, N_s)$ be the partition function in one sector with filling fractions $\{N_I\}$, and $Z(N'_1, \cdots, N_d')$ be the partition function corresponding to a different choice of filling fractions $\{N_I'\}$. In this case one can immediately write
\be\label{expinv}
{Z(N'_1, \ldots, N_d') \over Z(N_1, \ldots, N_d)} \sim \exp \biggl\{ -{1\over g_s} \sum_{I=1}^s (N_I-N_I') {\partial F_0(t_I) \over \partial t_I}\biggr\}.
\ee
\noindent
This expression means that, if we pick a set of filling fractions $(N_1, \ldots, N_s)$ as our reference background, \textit{all} the other sectors will \textit{not} be seen in $g_s$ perturbation theory. It is then only natural to regard them as different \textit{instanton sectors} of the matrix model.

Furthermore, notice that, depending on the value of the real part of the exponent appearing in (\ref{expinv}), the sectors with filling fractions $(N'_1, \cdots, N'_s)$ will be either exponentially suppressed or exponentially enhanced, with respect to the reference configuration $(N_1, \cdots, N_s)$. Let us consider for example a cubic potential where $N_1$ eigenvalues sit at the minimum of the potential and $N_2$ eigenvalues sit at the maximum. The sector with fillings $(N_1-1, N_2+1)$, which may be regarded as the one--instanton sector, will be more unstable than our reference configuration, and typically it will be exponentially suppressed. However, the anti--instanton sector with fillings $(N_1+1, N_2-1)$ is more stable and it will be exponentially enhanced. Therefore, in multi--cut matrix models (and in contrast to most field theories) instantons and anti--instantons have actions with opposite signs. As we shall see in the following, this leads to some subtleties.

Let us then consider a reference configuration characterized by arbitrary filling fractions $\{N_I\}$, $I=1, \ldots, s$, and regard any other set of filling fractions $\{N_I'\}$ as instanton configurations around our reference background. In this case, the determination of the instanton expansion of the partition function amounts to a Taylor expansion around the configuration with $t_I=g_s N_I$. For simplicity we will write down formulae for the two--cut case, and refer to \cite{eynard} for the formal, general expressions. We start from a reference configuration $(N_1, N_2)$ and write
\be\label{zexp}
Z = \sum_{\ell=-N_2}^{N_1} \zeta^\ell Z(N_1-\ell, N_2+\ell) = Z^{(0)} (N_1, N_2)  \sum_{\ell=-N_2}^{N_1} \zeta^\ell  Z^{(\ell)}, 
\ee
\noindent
where we have denoted $Z^{(0)}(N_1, N_2)=Z(N_1, N_2)$ and
\be
Z^{(\ell)}={Z(N_1-\ell,N_2+\ell) \over Z^{(0)}(N_1, N_2)}.
\ee
\noindent
Let us now assume that $N_I\not=0$ for $I=1, 2$. We recall that $t_I=g_s N_I$ and, as such, in terms of the 't~Hooft parameters one may equivalently write
\be
Z^{(\ell)}={Z(t_1-\ell g_s,t_2+\ell g_s ) \over Z^{(0)}(t_1, t_2)}.
\ee
\noindent
In the two--cut case it turns out to be rather convenient to use the variables
\be\label{st}
t=t_1 + t_2 \qquad \mathrm{and} \qquad s={1\over 2}( t_1 -t_2). 
\ee
\noindent
In this case, one immediately finds, by simply expanding around $g_s=0$ (or, equivalently, around $s=0$)
\be\label{zl}
Z^{(\ell)} = q^{\frac{\ell^2}{2}}\, \exp \left(- \frac{\ell A}{g_s}\right)\, \biggl\{ 1 - g_s \Bigl( \ell\, \partial_{s} F_1 + {\ell^3 \over 6}\, \partial_s^3 F_0 \Bigr) + \CO(g_s^2) \biggr\}.
\ee
\noindent
In this equation we have set
\be\label{aq}
A(t_I)=\partial_{s} F_0(t_I) \qquad \mathrm{and} \qquad q=\exp\Bigl( \partial_s^2 F_0 \Bigr).
\ee
\noindent
The first thing one recognizes is that $A$ corresponds to the action of an instanton obtained by eigenvalue tunneling from the first cut to the second cut. Notice that, due to (\ref{bmodel}), we can write $A$ as an integral over the spectral curve, obtaining the familiar result
\be
A=\int_{x_{2}}^{x_{3}}\rd x\, y(x). 
\ee

The expression (\ref{zl}) gives a general formula for the first terms in the $g_s$ expansion of the $\ell$--th instanton contribution to the partition function, in a general two--cut background. Notice however that, since $\ell$ can be either positive or negative, the expansion in (\ref{zexp}) will be a Laurent expansion in both $\xi=\exp(-A/g_s)$ and $\xi^{-1}$. In this case, the free energy will also be expressed as a Laurent series in $\xi$ and $\xi^{-1}$, but each coefficient in this series will be given by an infinite sum of terms. This is the subtlety we were previously referring to, and it comes from the fact that we are expanding around saddle--points which may have unstable directions.

There is a rather natural way to re--sum these terms by using theta--functions, as first introduced in \cite{bde, eynard}. At large $N_1$ and $N_2$ one can extend the sum in (\ref{zexp}) from $-\infty$ to $+\infty$, obtaining in this way
\be
Z = Z^{(0)}(t_1, t_2) \sum_{\ell=-\infty}^{+\infty} \zeta^\ell\, \xi^{\ell}\, q^{\frac{\ell^2}{2}}\, \biggl\{ 1 - g_s \Bigl( \ell\, \partial_{s} F_1 + {\ell^3 \over 6}\,  \partial_s^3 F_0 \Bigr) + \CO(g_s^2) \biggr\}.
\ee
\noindent
If one now exchanges the sum over $\ell$ with the expansion in $g_s$, we may immediately  write $Z$ in terms of the Jacobi theta--function
\be
\vartheta_3 (q|z) = \sum_{\ell=-\infty}^{+\infty} q^{\frac{\ell^2}{2}}\, z^{\ell},
\ee
\noindent
where $z \equiv \zeta\xi$, as
\be
Z = Z^{(0)}(t_1, t_2)\, \biggl\{  \vartheta_3(q|z) - g_s  \Bigl(  z\partial_z \vartheta_3(q|z)\, \partial_{s} F_1 + {1\over 6}\, (z\partial_z)^3 \vartheta_3(q|z)\, \partial^3_{s} F_0 \Bigr) + \CO(g_s^2) \biggr\}.
\ee
\noindent
As we shall explicitly show in the next section, this theta--function is well defined due to the fact that 
\be
\frac{\partial^2_s F_0}{2\pi\ri}
\ee
\noindent
is indeed the modulus of an elliptic curve, hence $|q|<1$. One may now use Jacobi's triple identity
\be
\vartheta_3(q|z) = \prod_{n=1}^{\infty}(1-q^n) \prod_{n=1}^{\infty}(1+ z q^{n-1/2}) \prod_{n=1}^{\infty}(1+z^{-1} q^{n-1/2})
\ee
\noindent
to write
\be
\log\, \vartheta_3(q|z) = \log \phi(q) + \sum_{\ell=1}^{\infty} {(-1)^{\ell} \over \ell}\, {z^\ell + z^{-\ell} \over q^{\ell\over 2} - q^{-{\ell\over 2}}},
\ee
\noindent
where 
\be
\phi(q) = \prod_{n=1}^{\infty}(1-q^n).
\ee
\noindent
The last two terms may also be written in terms of the quantum dilogarithm
\be
{\rm Li}_2^q (z) =\sum_{\ell=1}^{\infty} {(-1)^{\ell} \over \ell}\, {z^\ell \over q^{\ell\over 2} - q^{-{\ell\over 2}}}.
\ee
\noindent
This reorganization makes it possible to express the total free energy $F=\log Z$ in terms of an infinite series which formally has the structure of an instanton/anti--instanton expansion, as
\be
F = F^{(0)}(t_1, t_2) + \log \phi(q) + \sum_{\ell \not=0} {(-1)^{\ell} \over \ell (q^{\ell\over 2} - q^{-{\ell\over 2}})}\, \zeta^{\ell}\, \re^{-\ell A /g_s}\, \Bigl( 1 + \CO(g_s) \Bigr). 
\ee
\noindent
It is of course possible to write the $g_s$ corrections in a similar way. Notice that $\log \phi(q)$ gives a contribution to $F_1$ coming from instanton/anti--instanton interactions in the partition function.

In this paper we shall mostly explore another avenue. We will expand the multi--cut partition function (\ref{sumz}) around the most stable configuration, which generically is a one--cut configuration. In this way we will obtain general formulae for multi--instanton contributions in the one--cut matrix model by using multi--cut matrix models, generalizing the one--instanton formulae we first obtained in \cite{msw}. Before doing that, in the next section we shall give details on the solution of multi--cut matrix models which will then be useful in our subsequent computations.

\sectiono{Solving Multi--Cut Matrix Models}

In this section we will review the explicit solution of the multi--cut matrix model, focusing on the two--cut case. In particular, we will explicitly compute the several relevant quantities appearing in the multi--instanton formulae we have obtained in the previous section. An introduction to many of the results collected in the following may be found in, \textit{e.g.}, \cite{bde, akemann, leshouches}.

\subsection{General Results}

In order to solve for the density of eigenvalues (\ref{density}) we introduce, as in the one--cut case, the resolvent
\be\label{resolvent}
\omega (z) = \frac{1}{N} \left\langle \tr \frac{1}{z-M} \right\rangle = \frac{1}{N} \sum_{k=0}^{+\infty} \frac{1}{z^{k+1}} \left\langle \tr M^k \right\rangle,
\ee
\noindent
which has a standard genus expansion $\omega(z) = \sum_{g=0}^{+\infty} g_s^{2g}\, \omega_g (z)$ with
\be
\omega_0 (z) = \int \rmd\lambda\, \frac{\rho(\lambda)}{z-\lambda}.
\ee
\noindent
The normalization of the eigenvalue density 
\be
\int_{\CC}\rd \lambda\, \rho(\lambda)=1
\ee
\noindent
implies that 
\be\label{omas}
\omega_0(z) \sim \frac{1}{z}
\ee
\noindent
as $z \to + \infty$. Notice that the genus--zero resolvent determines the eigenvalue density as
\be
\rho(z) = - \frac{1}{2\pi\rmi} \left( \omega_0 (z+\rmi\epsilon) - \omega_0 (z-\rmi\epsilon) \right).
\ee
\noindent
One may compute $\omega_0(z)$ by making use of the large $N$ saddle--point equations 
of motion of the matrix model,
\be
\omega_0 (z+\rmi\epsilon) + \omega_0 (z-\rmi\epsilon) = \frac{1}{t} V'(z) = 2\, \mathsf{P} \int_{\CC} \rmd\lambda\, \frac{\rho(\lambda)}{z-\lambda}.
\ee
\noindent
For a generic multi--cut solution, the large $N$ resolvent is given by
\be\label{resf}
\omega_0(z) = \frac{1}{2t} \oint_{\CC} \frac{\rmd w}{2\pi\rmi}\, \frac{V'(w)}{z-w}\, {\sqrt{\sigma(z) \over \sigma(w)}},
\ee
\noindent
where we have introduced the notation
\be
\sigma(x) =\prod_{k=1}^{2s} (x-x_k).
\ee
\noindent
In (\ref{resf}), $\CC$ denotes a closed contour enclosing the union of intervals appearing in the standard multi--cut case, as we have discussed in the previous section. An equivalent way to describe the large $N$ solution is via the {\it spectral curve} $y(z)$, which is given by
\be
y(z) = V'(z) - 2t\, \omega_0(z) = M(z) \sqrt{\sigma(z)},
\ee
\noindent
where
\be\label{momentf}
M(z) = \oint_{\CC_{\infty}} \frac{\rmd w}{2\pi\rmi}\, {V'(w) \over (w-z) {\sqrt{\sigma(w)}}}
\ee
\noindent
is the moment function. The moments of $M(z)$, which are defined by \cite{akemann}
\be
M_i^{(k)} =\oint_{\CC} {\rd w \over 2\pi \ri}\, {V'(w) \over (w-x_i)^k {\sqrt{\sigma (w)}}},
\ee
can be easily calculated to be 
\be
M_i^{(k)}={1\over (k-1)!}\, {\rd^{k-1} \over \rd z^{k-1}}M(z)\Bigl|_{z=x_i} 
\ee
and will play an important role in the following. We shall also denote
\be
M_i\equiv M_i^{(1)} =M(x_i). 
\ee

In order to fully determine the large $N$ solution one still needs to specify the endpoints of the $s$ cuts, $\{x_k\}_{k=1,\ldots, 2s}$. The large $z$ asymptotics of the genus--zero resolvent immediately yield $s+1$ conditions for these $2s$ unknowns. They are
\be\label{asympcond}
\oint_{\CC} \frac{\rmd w}{2\pi\rmi}\, \frac{w^n V'(w)}{\sqrt{\sigma (w)}} = 2t\, \delta_{ns},
\ee
\noindent
for $n=0,1,\ldots,s$. In order to fully solve the problem, one still requires $s-1$ extra conditions. These are obtained by using (\ref{filfil}) and treating these filling fractions (or, equivalently, the partial `t~Hooft couplings (\ref{thooftmoduli})) as parameters, or moduli, of the model. Since $\sum_{I=1}^s \epsilon_I = 1$, in (\ref{filfil}) one indeed only finds $s-1$ conditions, as initially required.

Later on we shall need explicit expressions for the derivatives of the endpoints of the cuts, $x_j$, with respect to the 't~Hooft moduli, $t_j$. These are obtained as solutions to a linear system which we will now write down. Taking derivatives with respect to $t_j$ in (\ref{asympcond}), one immediately obtains 
\be\label{easyones}
\sum_{i=1}^{2s} M_{i}\, x_i^k\, \frac{\partial x_i}{\partial t_j} = 4\delta_{ks}, \quad k=0, \ldots, s. 
\ee
\noindent
For fixed $j$, this gives $s+1$ conditions for $2s$ quantities. The remaining $s-1$ conditions can be obtained precisely from the 't~Hooft moduli,
\be\label{deftj}
t_k = {1\over 2\pi} \int_{x_{2k-1}}^{x_{2k}} \rd \lambda\, M(\lambda) {\sqrt{|\sigma(\lambda)|}},
\ee
\noindent
by taking derivatives with respect to $t_j$. A similar calculation is done in \cite{akemann}, in its Appendix A. By deforming the contour away from infinity, we can write the moment function as
\be
M(\lambda) = -\oint_{\CC \cup \CC_{\lambda}} {\rd w \over 2 \pi \ri}\, {V'(w) \over (w-\lambda) {\sqrt {\sigma (w)}}},
\ee
\noindent
where $\CC_{\lambda}$ is a small contour around $w=\lambda$. Therefore, by taking derivatives with respect to $t_j$ in (\ref{deftj}), we obtain 
\be
\delta_{jk} = -{1\over 4\pi} \sum_{i=1}^{2s} {\partial  x_i \over \partial t_j}\int_{x_{2k-1}}^{x_{2k}} \rd \lambda\, \oint_{\CC \cup \CC_{\lambda}} {\rd \omega \over 2 \pi \ri} {V'(\omega) \over (\omega-\lambda)}\Bigl( {1\over \omega-x_i} - {1\over \lambda-x_i}\Bigr){\sqrt{\sigma(\lambda)} \over \sqrt{\sigma(\omega)}},
\ee
\noindent
where the derivative acted on $\sigma(\lambda)$ and $\sigma(\omega)$. Since
\be
{1\over \omega-x_i} - {1\over \lambda-x_i} = {\lambda - \omega \over (\omega-x_i)(\lambda-x_i)}, 
\ee
\noindent
we finally obtain
\be
\oint_{\CC \cup \CC_{\lambda}} {\rd \omega \over 2 \pi \ri}\, {V'(\omega) \over (\omega-\lambda)}\, \Bigl( {1\over \omega-x_i} -{1\over \lambda-x_i}\Bigr)\, {\sqrt{\sigma(\lambda)} \over \sqrt{\sigma(\omega)}} = - {\sqrt{\sigma(\lambda)} \over \lambda-x_i}\, M_i.
\ee
\noindent
Notice that the integrand no longer has a pole at $\omega=\lambda$, hence only the integral around $\CC$ contributes. We then end up with the following $s$ equations
\be
{1\over 4\pi} \sum_{i=1}^{2s} {\partial x_i \over \partial t_j}\, M_i \int_{x_{2k-1}}^{x_{2k}} \rd \lambda\, {\sqrt{\sigma(\lambda)} \over \lambda-x_i} = \delta_{jk}, \qquad k=1, \ldots, s.
\ee
\noindent
If we introduce the integrals \cite{akemann}
\be
K_{i,k} = \int_{x_{2k-1}}^{x_{2k}} \rd \lambda\, {{\sqrt{\sigma(\lambda)}} \over \lambda-x_i},
\ee
\noindent
we may then write
\be\label{difones}
{1\over 4 \pi} \sum_{i=1}^{2s} M_i\, K_{i,k}\, {\partial x_i \over \partial t_j} = \delta_{jk}, \qquad k=1, \ldots, s.
\ee
\noindent
These equations, together with (\ref{easyones}), fully determine the derivatives ${\partial x_i / \partial t_j}$. Observe that we have written down $2s+1$ equations in this procedure, but, as it turns out, only $2s$ of these equations are actually independent, uniquely determining our $2s$ unknowns. Other derivatives with respect to the 't~Hooft parameters can also be expressed in terms of the derivatives of the branch points. For example, from
\be
\frac{\partial M_i}{\partial x_i} = \frac{3}{2} M_i^{(2)} \qquad \mathrm{and} \qquad \frac{\partial M_i}{\partial x_k} = \frac{1}{2}\, \frac{M_i-M_k}{x_i-x_k}, \quad k\not=i,
\ee
\noindent
it is simple to find
\be\label{dermi}
{\partial M_i \over \partial t_j} = {3\over 2} M_i^{(2)} {\partial x_i \over \partial t_j} + {1\over 2} \sum_{k\not= i} {M_i-M_k \over x_i -x_k}  {\partial x_k \over \partial t_j}.
\ee

\subsection{The Two--Cut Matrix Model} 

In the two--cut case one can write very concrete results for the different quantities involved in the above solution, in terms of elliptic functions \cite{akemann, bde, kmt}. We now review some of these results.

The support of the density of eigenvalues is given in this case by
\be
A_1 \cup A_2 = [x_1, x_2] \cup [x_3, x_4].
\ee
\noindent
We define, as in \cite{bde},
\begin{equation}\label{Kdef}
\CK = \int_{x_1}^{x_2} {\rd {z}\over\sqrt{|\sigma(z)|}} \,=\, {2\over \sqrt{{(x_1-x_3)(x_2-x_4)}}}\, K(k), \qquad  k^2 = {(x_1-x_2)(x_3-x_4)\over(x_1-x_3)(x_2-x_4)},
\end{equation}
\noindent
where $K(k)$ is the complete elliptic integral of the first kind. Similarly, 
\begin{equation}\label{Kpdef}
\CK' = \int_{x_2}^{x_3} {\rd {z}\over\sqrt{|\sigma(z)|}} = {2\over\sqrt{(x_1-x_3)(x_2-x_4)}}\, K(k'), \quad k'^2 = 1-k^2 = {(x_4-x_1)(x_3-x_2)\over(x_1-x_3)(x_2-x_4) }.
\end{equation}
\noindent
We refer the reader to the appendix for a list of definitions and conventions concerning elliptic functions. One further defines the modular parameter $\tau$ of the elliptic curve $y^2=\sigma(x)$ by
\begin{equation}\label{eq:tauK}
\tau\, = \ri {\CK' \over \CK}\, = \ri {K(k')\over K(k)}.
\end{equation}
\noindent
It is now easy to show that the second derivative appearing in the one--loop coefficient of the $\ell$--instanton sector, (\ref{aq}), is given by
\be\label{secondtau}
{\partial^2 F_0 \over \partial s^2}=2\pi \ri \tau. 
\ee
\noindent
To do this, we adapt an argument from \cite{bde}. From the special geometry relation (\ref{bmodel}) it follows that
\be
{\partial F_0 \over \partial s} = \int_{x_2}^{x_3} \rd x\, (V'(x) - 2t \omega_0(x)), 
\ee
\noindent
hence
\be\label{toplug}
{\partial^2 F_0 \over \partial s^2} = -2 \int_{x_2}^{x_3} \rd x\, {\partial (t \omega_0(x)) \over \partial s}. 
\ee
\noindent
Using the asymptotic behavior of $\omega_0(x)$ in (\ref{omas}) it is easy to prove that
\be
{\partial (t \omega_0(x)) \over \partial s} = {C(s,t) \over {\sqrt{\sigma(x)}}}, 
\ee
\noindent
where $C(s,t)$ does not depend on $x$. Its value can be fixed by using that
\be
t_I = \oint_{A_I} {\rd x \over 2 \pi \ri}\, t\omega_0(x). 
\ee
\noindent
Indeed, since
\be
{\partial t_1 \over \partial s} = 1 = \oint_{\CC_1} {\rd x \over 2 \pi \ri}\, {C(s,t) \over  {\sqrt{\sigma(x)}}},
\ee
\noindent
it follows that
\be
C(s,t) = -{ \pi \over {\CK}}. 
\ee
\noindent
Now plugging this result inside (\ref{toplug}), we immediately find (\ref{secondtau}).  
 
In the two--cut case, the equations for $\partial x_i /\partial t_j$ which were obtained in the previous section, can also be explicitly written in terms of elliptic functions. The required integrals $K_{i,k}$ were evaluated for all $i=1, \ldots, 4$, and for $k=2$, in Appendix B of \cite{kmt}, and we can use the expressions therein to solve the resulting system of equations. One may also use a procedure identical to the one in \cite{kmt} to obtain the $K_{i,k}$ integrals when $k=1$, with the same final result for the $\partial x_i /\partial t_j$ derivatives. In particular, one finds the resulting simple answer
\be\label{dxi}
{\partial x_i \over \partial s} = {4 \pi \over M_i\, \CK}\, {1\over \prod_{j\not=i}(x_i-x_j)}.
\ee

Using the above formulae, it is easy to obtain explicit expressions for the other derivatives of the free energy appearing in (\ref{zl}), in particular the two--loop contribution to the $\ell$--instanton sector. The cubic derivative $\partial_s^3 F_0$ can be obtained directly from (\ref{secondtau}) as
\be
\partial_s^3 F_0 = 2\pi \ri\, {\partial \tau \over \partial k^2}\, {\partial k^2 \over \partial s}. 
\ee
\noindent
The first factor in the right hand side may be computed with the use of the identities
\be
\ba
{\rd K \over \rd k^2} &= {1\over 2 k^2 k'^2} (E -k'^2 K), \\
{\rd K' \over \rd k^2} &= -{1\over 2 k^2 k'^2} (E' -k^2 K'), 
\ea
\ee
\noindent
while the second factor is more involved and can be obtained from (\ref{dxi}). The result, for a generic two--cut matrix model, is
\bea
\frac{\partial k^2}{\partial s} &=& \frac{2 \pi}{M_1 \cdots M_4\, K(k)}\, \frac{1}{(x_1-x_3)^{3/2} (x_2-x_4)^{3/2}}\, \frac{1}{\prod_{i<j} (x_i-x_j)} \cdot \nonumber \\
&& \cdot \left( M_1 M_2 M_3\, (x_1-x_2)^2 (x_1-x_3)^2 (x_2-x_3)^2 + \right. \nonumber \\
&& + M_1 M_2 M_4\, (x_1-x_2)^2 (x_1-x_4)^2 (x_2-x_4)^2 + M_1 M_3 M_4\, (x_1-x_3)^2(x_1-x_4)^2 (x_3-x_4)^2 + \nonumber \\
&& \left. + M_2 M_3 M_4\, (x_2-x_3)^2 (x_2-x_4)^2 (x_3-x_4)^2 \right).
\eea
\noindent
Putting everything together we finally obtain
\bea\label{fullthreef}
\frac{1}{6}\, \frac{\partial^3 F_0}{\partial s^3} &=& \frac{\pi^3}{6\, M_1 \cdots M_4\, K^3(k)}\, \frac{\sqrt{(x_1-x_3)(x_2-x_4)}}{(x_1-x_2)(x_2-x_3)(x_1-x_4)(x_3-x_4)}\, \frac{1}{\prod_{i<j} (x_i-x_j)} \cdot \nonumber \\
&& \cdot \left( M_1 M_2 M_3\, (x_1-x_2)^2 (x_1-x_3)^2 (x_2-x_3)^2 + \right. \nonumber \\
&& + M_1 M_2 M_4\, (x_1-x_2)^2 (x_1-x_4)^2 (x_2-x_4)^2 + M_1 M_3 M_4\, (x_1-x_3)^2(x_1-x_4)^2 (x_3-x_4)^2 + \nonumber \\
&& \left. + M_2 M_3 M_4\, (x_2-x_3)^2(x_2-x_4)^2 (x_3-x_4)^2 \right).
\eea

Another quantity which can be computed in the two--cut case in terms of elliptic functions is $F_1(t_1, t_2)$ \cite{akemann,kmt}. It is given by
\be\label{Fonematrix}
F^{(1)}=-{1\over24}\sum_{i=1}^4\ln M_i-{1\over2}\ln K(k) -{1\over12}\sum_{i<j}\ln(x_i-x_j)^2 +{1\over8}\ln(x_1-x_3)^2+{1\over8}\ln(x_2-x_4)^2.
\ee
\noindent
This is, in fact, the last ingredient required in order to compute the complete two--loop coefficient in the $\ell$--instanton sector expansion, where we are still missing the expression for $\partial_s F_1$ in the two--cut case. As it turns out, the resulting expression is rather long, and not particularly illuminating and, as such, we present it in the appendix.

\sectiono{Multi--Instantons in the One--Cut Matrix Model}

We shall now shift gears and consider multi--instantons in the one--cut matrix model as a particular, degenerate case of our above two--cut considerations; in this way generalizing the one--instanton results we have previously obtained in \cite{msw}.

\subsection{General Formulae}

As we mentioned at the end of section 2, one of our goals in this paper is to focus on the expansion of the multi--cut partition function around its most stable cut, suppressing in this way the anti--instanton configurations. This is the natural choice, for example, if we are interested in computing a convergent matrix integral in terms of a perturbative series in $g_s$ plus exponentially small corrections, as in \cite{mmnp}. We will assume that we are in a generic situation and the potential has a unique global minimum (like in \figref{cubictunnel}). If we call $A_1$ the cut surrounding this minimum, the most stable configuration has filling fractions
\be\label{refcon}
(N,0, \cdots, 0)
\ee
\noindent
and it is a one--cut solution of the model. Any other configuration will be exponentially suppressed with respect to this one, and a convergent multi--cut matrix integral can be computed at large $N$ by considering the partition function $Z(N, 0, \cdots, 0)$ and then adding exponentially suppressed contributions from the other configurations. In this way we obtain a multi--cut configuration as a sum of multi--instantons in the one--cut matrix model. In particular, this produces general formulae for the multi--instantons amplitudes of the one--cut solution.

We will focus for simplicity on stable one--cut configurations which arise as limiting cases of two--cut matrix models. This means that we will only consider one type of instanton, arising from eigenvalue tunneling from the filled cut to the empty cut. The generalization to models with more than two cuts should be completely straightforward\footnote{The situation we analyze in here is similar to the ``birth of a cut'' transition studied in \cite{eynardbirth}. For previous work on phase transitions of this type in the Hermitian matrix model see, \textit{e.g.}, \cite{jurkiewicz, mandal, david1}.}.

\FIGURE[ht]{
\leavevmode
\centering
\epsfysize=3.5cm
\hspace{3cm}
\epsfbox{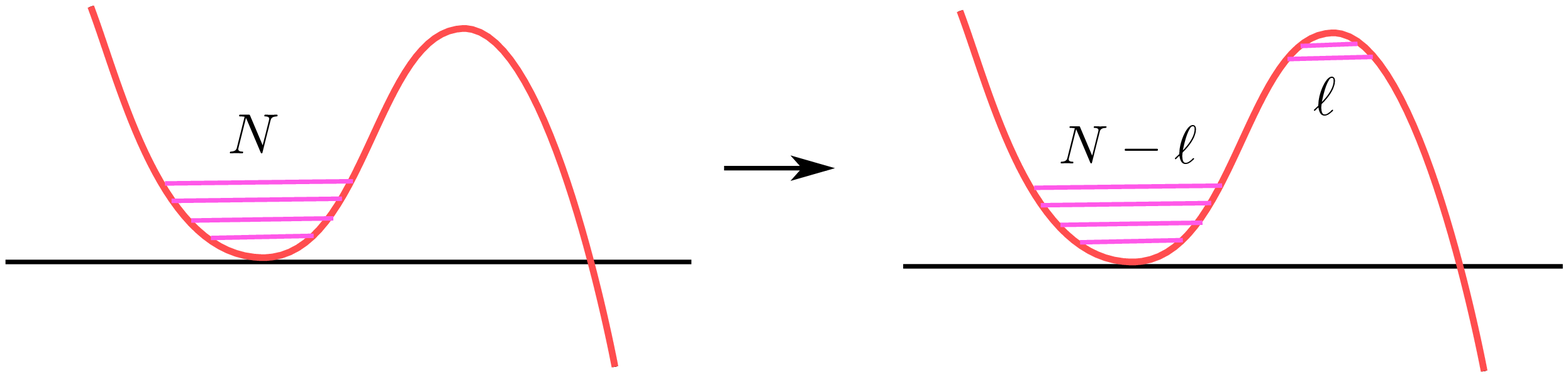}
\hspace{3cm}
\caption{The total partition function (\ref{znaif}) is a sum over all possible instanton sectors. The $\ell$--th instanton sector is obtained by tunneling $\ell$ eigenvalues from the most stable, one--cut configuration, to the other (unstable) saddle--point.}
\label{cubictunnel}
}

The total partition function is obtained by summing over all possible numbers $\ell$ of eigenvalues  tunneling from the filled cut to the empty one, as
\be\label{znaif}
Z = \sum_{\ell= 0}^N \zeta^\ell\, Z(N-\ell, \ell).
\ee
\noindent
We are interested in the 't~Hooft limit in which $N\rightarrow \infty$ with $t =g_s N$ fixed, and $\ell \ll N$. From the point of view of the spectral curve, the partition function $Z(N-\ell, \ell)$ corresponds to a two--cut curve in which one cut is filled with $N-\ell$ eigenvalues, with $N \gg 1$, and the other cut is nearly but not completely pinched, since it contains $\ell \ll N$ eigenvalues (see \figref{multinst}). In the 't~Hooft limit the finite sum in (\ref{znaif}) becomes an infinite one, and one can write
\be
Z= Z^{(0)}(t)\, \biggl\{ 1 + \sum_{\ell=1}^\infty \zeta^\ell\, Z^{(\ell)} \biggr\}, 
\ee
\noindent
where $Z^{(0)}(t)=Z(N,0)$ is the total partition function of the one--cut matrix model, \textit{i.e.}, the model in which the $N$ eigenvalues sit at the minimum, and 
\be
Z^{(\ell)} = {Z(t - \ell g_s, \ell g_s) \over Z^{(0)}}. 
\ee
\noindent
Just like we did before, we can try to evaluate $Z^{(\ell)}$ by expanding the numerator around $g_s=0$. If one tries this there is a subtlety, however, since the free energies $F_g(t_1, t_2)$ are \textit{not} analytic at $t_2=0$. Geometrically, this corresponds to the fact that we are expanding around a configuration in which the second cut of the spectral curve is completely pinched. To understand the origin of this nonanalyticity, write  
\be
F_g(t_1, t_2) = F^{\rm G}_g(t_2) + \widehat F_g(t_1, t_2), 
\ee
\noindent
where $F^{\rm G}_g(t)$ are the genus $g$ free energies of the gauged Gaussian matrix model with 't~Hooft parameter $t$, \textit{i.e.}, 
\be
F_0^{\rm G}(t) = {1\over 2} t^2\, \Bigl( \log t - {3\over 2} \Bigr), \qquad F_1^{\rm G}(t) = - {1\over 12} \log t, \qquad \cdots
\ee
\noindent
It is simple to see that $\widehat F_g (t_1, t_2)$ is indeed {\it analytic} at $t_2=0$, so that the lack of analyticity of the matrix model free energy at $t_2=0$ is due to the Gaussian contribution $F^{\rm G}(t_2)$. This issue has been clarified in \cite{ov} in a slightly different context: the Gaussian part of the matrix model free energy comes from the measure, and it is not analytic when the 't~Hooft parameter vanishes. The ``regularized'' $\widehat F_g(t_1, t_2)$ comes from re--summing the perturbation theory double--line diagrams with genus $g$ and it is an analytic function.

\FIGURE[ht]{
\leavevmode
\centering
\epsfysize=4cm
\hspace{3cm}
\epsfbox{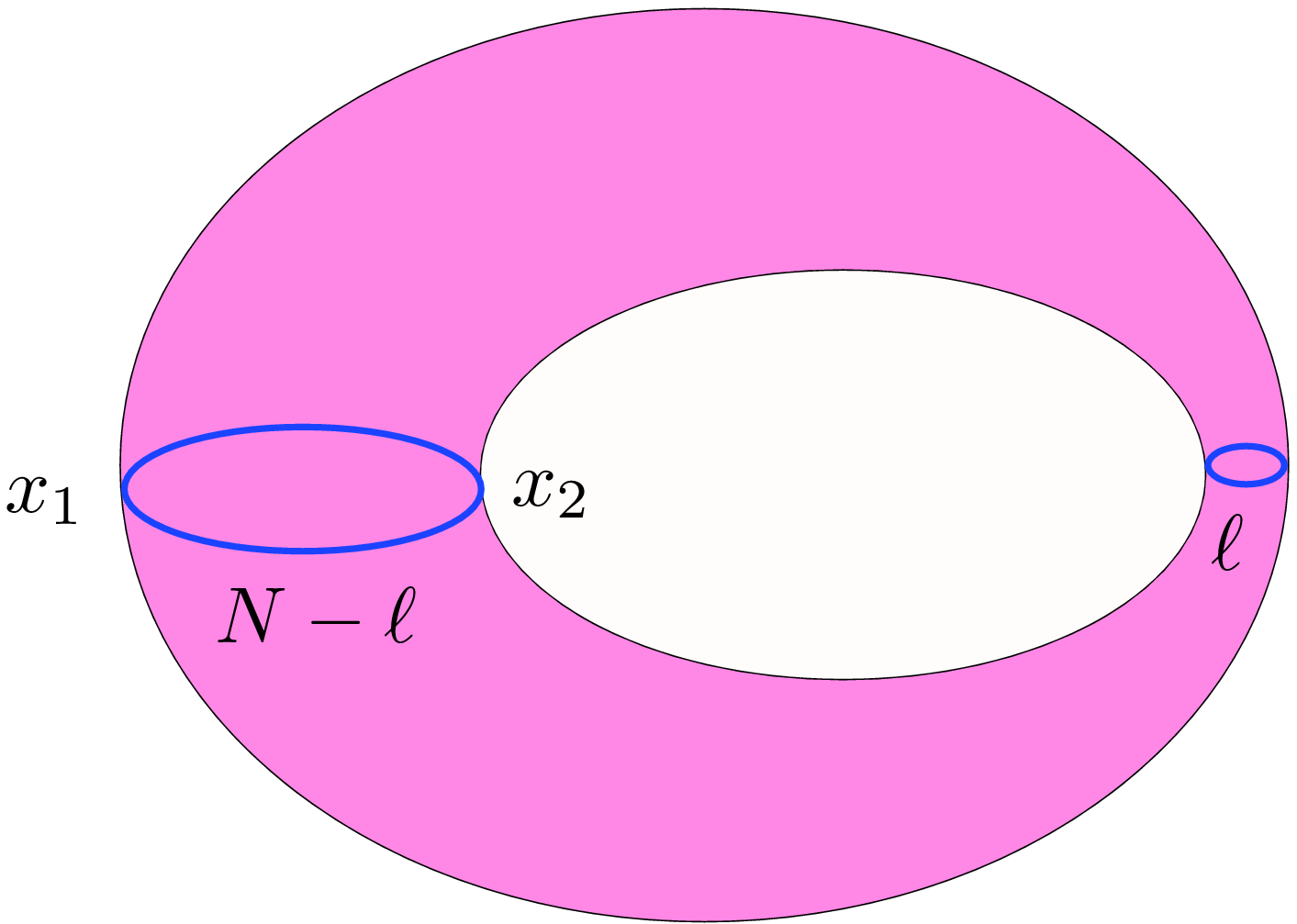}
\hspace{3cm}
\caption{The partition function in the $\ell$--th instanton sector corresponds to a spectral curve in which one of the cuts is filled with $N-\ell$ eigenvalues, with $N \gg 1$, while the other cut is nearly pinched, since it contains $\ell \ll N$ eigenvalues.}
\label{multinst}
}

Physically, the reason for the appearance of this singularity is that in this problem $t_2=\ell g_s$ and $\ell$ is small as compared to $N$. Therefore, it is not appropriate to treat the integration over the $\ell$ eigenvalues from the point of view of the large $N$ expansion. Instead, they should be integrated exactly. This argument also suggests that, in order to regularize the computation, we should subtract $F^{\rm G}(t_2)$ from the total free energy and at the same time multiply $Z^{(\ell)}$ by the {\it exact} partition function $Z^{\rm G}_\ell$ for the gauged Gaussian matrix model with $\ell$ eigenvalues,
\be
Z^{\rm G}_\ell = {g_s^{\ell^2/2} \over (2\pi)^{\ell/2}}\, G_2(\ell+1),
\ee
\noindent
where $G_2(\ell+1)$ is the Barnes function
\be
G_2(\ell+1)=\prod_{i=0}^{\ell-1} i!\, .
\ee
\noindent
The appropriate expression for the partition function around the $\ell$--instanton configuration is then
\be\label{interzn}
Z^{(\ell)} = Z^{\rm G}_\ell\, \exp \biggl[ \sum_{g\ge 0} g_s^{2g-2} \left( \widehat F_g (t-\ell g_s, \ell g_s) - F_g (t) \right)\biggr].
\ee
\noindent
We can now expand the exponent appearing in (\ref{interzn}) around $g_s=0$, since it is analytic, and we obtain, up to two loops, 
\be\label{prediction}
Z^{(\ell)} = {g_s^{\ell^2/2} \over (2\pi)^{\ell/2}}\, G_2(\ell+1)\, \zeta^\ell\, \hat q^{\frac{\ell^2}{2}}\, \exp \left( - \frac{\ell \widehat A}{g_s} \right) \biggl\{ 1 - g_s \Bigl( \ell\, \partial_{s} \widehat F_1 (t) + {\ell^3 \over 6}\, \partial_s^3 \widehat F_0 (t) \Bigr) + \CO(g_s^2)\biggr\}.
\ee
\noindent
In this equation 
\be\label{hata}
\widehat A(t) = \partial_{s} \widehat F_0 \qquad \mathrm{and} \qquad \hat q = \exp \Bigl(  \partial_s^2 \widehat F_0 \Bigr).
\ee
\noindent
Notice that all derivatives appearing in both (\ref{prediction}) and (\ref{hata}) are evaluated at $t_1=t$ and $t_2=0$.

We can now compute the free energy as
\be
F=\log Z=\sum_{\ell=0}^{\infty} z^{\ell}\, F^{(\ell)}
\ee
\noindent
where we set
\be
z= \zeta\, \re^{-\widehat A/g_s}.
\ee
\noindent
Since we are expanding around the most stable configuration $Z$ now has a Taylor series in $z$, and $F$ will be well--defined as a formal series in positive powers of $z$. This series is appropriately written in terms of ``connected'' contributions
\be
\log \biggl\{ 1 + \sum_{\ell = 1}^{+\infty} \zeta^\ell\, Z^{(\ell)} \biggr\} = \sum_{\ell = 1}^{\infty} \zeta^\ell\, Z^{(\ell)}_{(c)}, 
\ee
\noindent
where
\be
Z^{(\ell)}_{(c)} = \sum_{s\ge 1} {(-1)^{s-1}\over s} \sum_{k_1 + \cdots + k_s=\ell} Z^{(k_1)} \cdots Z^{(k_s)} = Z^{(\ell)} - {1\over 2} \sum_{k=1}^{\ell-1} Z^{(\ell-k)} Z^{(k)} + \cdots. 
\ee
\noindent
We then deduce that the $\ell$--instanton contribution to the free energy of a one--cut matrix model is given by 
\be
z^{\ell}\, F^{(\ell)} = \zeta^\ell\, Z^{(\ell)}_{(c)}. 
\ee
\noindent
The explicit expression of $F^{(\ell)}$ at two loops is
\be\label{multiformula}
\ba 
F^{(1)} &= {g_s^{1/2} \over {\sqrt{2\pi}}}\, \hat q^{1/2}\, \biggl\{ 1 - g_s \Bigl( \partial_s \widehat F_1 + {1 \over 6} \partial_s^3 \widehat F_0\Bigr) + \CO(g^2_s) \biggr\}, \\
F^{(\ell)} &={(-1)^{\ell-1} g_s^{\ell/2} \over \ell\, (2\pi)^{\ell \over 2}}\, \hat q^{\ell/2}\, \biggl\{  1- \ell\, g_s \Bigl( \partial_s \widehat F_1 + {1 \over 6} \partial_s^3 \widehat F_0 + \hat q \Bigr) + \CO(g^2_s) \biggr\}, \qquad \ell\ge 2,
\ea
\ee
\noindent
yielding formulae for the two--loops $\ell$--instanton amplitude in an arbitrary one--cut matrix model. 

\subsection{The Nature of the Instanton Gas}

An interesting point to notice is that our result for $Z^{(\ell)}$, (\ref{prediction}), is {\it not} typical of a dilute instanton gas. Indeed, in a dilute instanton gas one has that, at leading order in the coupling constant,
\be
Z^{(\ell)} \approx {1\over \ell!} \left( Z^{(1)} \right)^\ell. 
\ee
\noindent
If this was the case here $Z^{(\ell)}$ would scale as $g_s^{\ell/2}$, but it is clear from (\ref{prediction}) that this is not the case: $Z^{(\ell)}$ scales as $g_s^{\ell^2/2}$. The reason why this happens is due to the presence of the Vandermonde determinant, and it can be easily understood from a simple scaling argument. The $\ell$--instanton integral is roughly of the form 
\be
Z^{(\ell)} \approx \int \prod_{i=1}^\ell \rd \lambda_i\, \Delta^2 (\lambda) \exp \biggl[ - {1\over 2g_s} V_{\rm eff}''(x_0) \sum_{i=1}^\ell (\lambda_i-x_0)^2 \biggr].
\ee
\noindent
If we change variables to $u_i=(\lambda_i-x_0)/g^{1/ 2}_s$ it immediately follows that the measure for the $\ell$ eigenvalues leads to the factor $g_s^{\ell/2}$ typical of a dilute instanton gas. However, in the same way the Vandermonde determinant leads to an extra power of $g_s^{\ell(\ell-1)/2}$. The instanton gas in a matrix model should be rather regarded as an \textit{ultra--dilute} instanton gas since, at weak coupling $g_s \ll 1$, the partition function for $\ell$ instantons is even more suppressed than in the usual instanton gas. Physically, the ultra--diluteness is of course due to the eigenvalue repulsion.

In particular, we disagree with the analysis of multi--instantons in the one--cut matrix model proposed in section 2 of \cite{lvm}. In that paper, the Vandermonde interaction between the instantons is set to one and the resulting integral factorized. This cannot be done without jeopardizing the very scaling of $Z^{(\ell)}$ with $g_s$. We shall later on verify this scaling, as well as other details of our main formulae (\ref{prediction}) and (\ref{multiformula}), against other techniques for computing multi--instanton effects.

\subsection{Explicit Calculations}

Our main multi--instanton formulae (\ref{prediction}) and (\ref{multiformula}) involve various derivatives of two--cut free energies, in the limit in which one of the cuts shrinks down to zero size, \textit{i.e.}, in the $t_2\rightarrow 0$ limit. We will now derive explicit expressions for these derivatives and, in particular, we shall check that the expression obtained for $F^{(1)}$ agrees with the result we previously obtained in \cite{msw}.

The limit $t_2\rightarrow 0$ corresponds geometrically to a degeneration of the curve in which the $A_2$ cycle shrinks to zero size, \textit{i.e.}, $x_3\rightarrow x_4 \equiv x_0$. The spectral curve of the two--cut problem becomes
\be
y(x) \rightarrow M(x) (x-x_0) {\sqrt{(x-x_1)(x-x_2)}}.
\ee
\noindent
Notice that here $M(x)$ is the moment function of the original two--cut problem. The moment function of the one--cut problem, obtained by degeneration, is then given by\footnote{Recall that, in the one--cut problem, it is precisely the requirement $M_{1} (x_0) = 0$ that allows for the nontrivial saddle--point $x_0$ characterizing the one--instanton sector \cite{msw}.}
\be\label{momentrel}
M_1(x) = M(x)(x-x_0). 
\ee

We can now compute the quantities appearing in the $\ell$--instanton free energies (\ref{multiformula}). First of all, the instanton action is given by
\be
\widehat A(t) = \lim_{t_2\to 0} \Bigl( \partial_s F_0(t, t_2) - \partial_{t_2} F_0^{\rm G}(t_2) \Bigr). 
\ee
\noindent
Both quantities appearing here are regular at $t_2=0$ and, from (\ref{bmodel}), one immediately finds
\be
\widehat A =\int_{x_2}^{x_0} \rd x\, M_1(x) {\sqrt{(x-x_1)(x-x_2)}}, 
\ee
\noindent
which is the instanton action in the one--cut case. 

We next compute
\be
{1\over 2} \partial_s^2 \widehat F_0(t,0) = {1\over 2} \lim_{t_2\to 0} \Bigl( \partial^2_s F_0(t, t_2) - \partial^2_{t_2} F_0^{\rm G}(t_2) \Bigr) = \lim_{x_4\rightarrow x_3} \biggl( \pi \ri \tau - {1\over 2} \log t_2 \biggr). 
\ee
\noindent
In the limit $x_4 \rightarrow x_3$ the elliptic modulus appearing in (\ref{Kdef}) vanishes and both $\tau$ and $\log t_2$ diverge. In order to calculate this limit, as well as similar ones, we will set
\be
\epsilon=x_4-x_3. 
\ee
\noindent
As $\epsilon \rightarrow 0$ the leading behavior of the elliptic modulus is given by
\be\label{leadel}
\pi \ri \tau \sim \log {k^2 \over 16}, 
\ee
\noindent
\textit{i.e.}, it diverges as $\log \epsilon$. According to our general argument above, this divergence should be removed by subtracting $(\log t_2)/2$. In an appendix we compute the expansion of $t_2$ in powers of $\epsilon$. In particular, we find that at leading order $t_2 \sim \epsilon^2$, so indeed the cancellation takes place. Putting together (\ref{leadel}) and (\ref{epstwo}) we find
\be\label{result}
\hat q^{1/2}= \lim_{\epsilon \to 0}\, {k^2 \over 16 {\sqrt{t_2}}} = {x_1-x_2\over 4} {1\over {\sqrt{M(x_0) [(x_1-x_0)(x_2-x_0)]^{5\over 2}}}}.
\ee
\noindent
Since
\be
M(x_0)=M_1'(x_0)
\ee
\noindent
(\ref{result}) is in complete agreement with the formula obtained in \cite{msw} for the one--loop contribution to $F^{(1)}$ (the factor $1/{\sqrt{2\pi}}$ appearing in the formula of \cite{msw} is already included in (\ref{multiformula})).
 
We now proceed with the calculation of the quantities appearing at two loops. We shall first calculate $\partial_s^3 \widehat F_0$, which is defined by
\be
{1\over 6} \partial_s^3 \widehat F_0(t,0) = \lim_{\epsilon \to 0} \biggl( {1 \over 6} \partial_s^3  F_0(t,t_2) - {1\over 6 t_2} \biggr)
\ee
\noindent
and should be regular. Indeed, it is easy to see that (\ref{fullthreef}) has a pole of order two in $\epsilon=x_4-x_3$. Making use of (\ref{epstwo}) one can explicitly check that both the divergent pieces in $\epsilon^{-2}$ and $\epsilon^{-1}$ are removed by subtracting $1/(6 t_2)$, and we finally get
\be\label{zerocubicder}
\ba
{1\over 6} \partial_s^3 \widehat F_0(t,0) &= {1\over {\sqrt{(x_0-x_1)(x_0-x_2)}}}\, \biggl\{ \frac{8 (x_0-x_1)^2}{(x_1-x_2)^2 (x_0-x_2)^2\, M(x_2)} + \frac{8 (x_0-x_2)^2}{(x_1-x_2)^2 (x_0-x_1)^2\, M(x_1)} \\
& + \frac{4 M'(x_0)^2}{M(x_0)^3} - \frac{3 M''(x_0)}{2 M(x_0)^2} - \frac{17 (x_1+x_2-2 x_0) M'(x_0)}{2 (x_0-x_1) (x_0-x_2)\, M(x_0)^2} + \\
& + \frac{77 x_1^2 + 118 x_1 x_2 + 77 x_2^2 - 272 x_0 \left(x_1 + x_2 - x_0\right)}{8 (x_0-x_1)^2  (x_0-x_2)^2\, M(x_0)} \biggr\}.
\ea
\ee

A similar computation can be done in the case of $F_1$. Again, from the explicit expression (\ref{dsf12cut}) it is easy to see that  $\partial_s F_1$ is singular as $\epsilon \rightarrow 0$, but 
\be
\partial_s \widehat F_1(t,0) = \lim_{\epsilon \to 0} \biggl( \partial_s  F_1(t,t_2) + {1\over 12 t_2}\biggr)
\ee
\noindent
turns out to be regular, as expected, and can be computed to be 
\be
\ba\label{onecubicder}
\partial_s \widehat F_1(t,0) &= {1\over {\sqrt{(x_0-x_1)(x_0-x_2)}}}\, \biggl\{ \frac{M'(x_0)^2}{6 M(x_0)^3} - \frac{M''(x_0)}{8 M(x_0)^2} - \frac{(x_1+x_2-2 x_0) M'(x_0)}{24 (x_0-x_1) (x_0-x_2)\, M(x_0)^2} + \\
& + \frac{19 (x_1^2+x_2^2) - 22 x_1 x_2 - 16 x_0 \left(x_1+x_2-x_0 \right)}{96 (x_0-x_1)^2 (x_2-x_0)^2 M(x_0)} - \frac{(x_0-x_2) (x_2 - 3 x_1 + 2 x_0)}{3 (x_1-x_2)^2 (x_0-x_1)^2 M(x_1)} - \\
& - \frac{(x_0-x_1) (x_1 - 3 x_2 + 2 x_0)}{3 (x_1-x_2)^2 (x_0-x_2)^2 M(x_2)} - \frac{(x_0-x_2)\, M'(x_1)}{4 (x_1-x_2) (x_0-x_1)\, M(x_1)^2} + \\
& + \frac{(x_0-x_1)\, M'(x_2)}{4 (x_1-x_2) (x_0-x_2)\, M(x_2)^2}  \biggr\}.
\ea
\ee
\noindent
This result, together with (\ref{zerocubicder}), gives a completely explicit result for the amplitudes (\ref{prediction}) and (\ref{multiformula}) in terms of geometrical data of the spectral curve for the the one--cut matrix model.

From the above expressions one easily checks that 
\be
{1\over 6} \partial_s^3 \widehat F_0(t,0) + \partial_s \widehat F_1(t,0)
\ee
\noindent
agrees with the two--loop result in equation (3.63) of \cite{msw}, after taking into account that $x_1, x_2$ were denoted by $a,b$ in there, and that the moment functions are related by (\ref{momentrel}) (in an appendix, we include some further relations between the moment functions of the one and two--cut models, in order to make the comparison more explicit). This tests our expression (\ref{multiformula}) for $F^{(1)}$ at two loops against our calculation of the one--instanton amplitude in \cite{msw}, with complete agreement.

\subsection{Multi--Instantons in the Cubic Matrix Model}

Our results (\ref{prediction}) and (\ref{multiformula}) provide general expressions for multi--instanton amplitudes in one--cut matrix models. We shall now exemplify and test these formulae, within the multi--instanton set--up, in the cubic matrix model. To perform the test, we have to compute multi--instanton amplitudes independently and, in order to accomplish this, we will use the trans--series formalism based on orthogonal polynomials which was recently developed in \cite{mmnp}.

Let us briefly summarize the formalism of \cite{mmnp}, to which we refer the reader for further details and examples. We first recall that the orthogonal polynomials $p_n(\lambda)$ for the potential $V(\lambda)$ are defined by
\be
\int {\rd\lambda\over 2\pi}\, \re^{-{V(\lambda)\over g_s}}\, p_n(\lambda) p_m(\lambda)= h_n\, \delta_{nm}, \qquad n>0,
\ee
\noindent
where $p_n$ are normalized by requiring that $p_n\sim \lambda^n+\cdots$. It is well known (see for example \cite{dfgz}) that the partition function \eqref{pf} of the matrix model can be expressed as
\be
Z_N=\prod_{i=0}^{N-1}h_i=h_0^N\, \prod_{i=1}^N r_i^{N-i}.
\ee
\noindent
The coefficients 
\be
r_n={h_n\over h_{n-1}}
\ee
\noindent
satisfy recursion relations depending on the shape of the potential. They also obviously satisfy
\be\label{rZ}
r_n={Z_{n-1}Z_{n+1}\over Z_n^2}.
\ee 
\noindent
In the limit $N\rightarrow \infty$, $\frac{g_s n}{N}$ becomes a continuous variable that we will denote by $z$, and $r_n$ is promoted to a function, $R(z,g_s)$, for which we have the multi--instanton trans--series ans\"atz
\be\label{Rexpansion}
R(z,g_s)=\sum_{\ell=0}^\infty C^\ell R^{(\ell)}(z,g_s)\, \re^{-\ell A(z)/g_s},
\ee
\noindent
where the $\ell$--instanton coefficient can be written as 
\be\label{Rl}
R^{(\ell)}(z,g_s)=R^{(\ell)}_0(z)\left(1+\sum_{n=1}^{\infty}g_s^n\, R^{(\ell)}_n(z)\right)
\ee
\noindent
and $C$ is a nonperturbative ambiguity. The coefficients $R^{(\ell)}_n(z,g_s)$ can be computed from the so--called pre--string equation, a difference equation which may be derived as the continuum limit of the recursion relations satisfied by the coefficients $r_n$. For any polynomial potential the pre--string equation can be written down explicitly \cite{dfgz}. Furthermore, once \eqref{Rexpansion} has been found, we can then extract the free energy $F(z,g_s)=\log Z$ from the continuum limit of expression \eqref{rZ}, namely
\be\label{FlogR}
F(z+g_s,g_s)+F(z-g_s,g_s)-2F(z,g_s)=\log R(z,g_s).
\ee
\noindent
The total free energy $F(z,g_s)$ also has an instanton trans--series expansion of the form
\be
F(z,g_s) = \sum_{\ell \ge 0} \re^{-\ell A (z)/ g_s} F^{(\ell)}(z,g_s), 
\ee
\noindent
where 
\be
F^{(\ell)}(z,g_s)=F^{(\ell)}_0(z)\left(1+\sum_{n=1}^{\infty}g_s^n\, F^{(\ell)}_n(z)\right), \quad \ell \ge 1. 
\ee
\noindent
The different coefficients $F^{(\ell)}_n(z)$ can be easily obtained from $A(z)$ and $R^{(\ell)}_n(z,g_s)$, see \cite{mmnp} for some explicit expressions. Notice that in here the nonperturbative ambiguity $C$ is an integration constant and it cannot be fixed within this method. In particular, it might differ from the nonperturbative ambiguity $\zeta$ appearing in the partition function (\ref{sumz}) by a multiplicative factor independent of $z$, the 't~Hooft parameter (see \cite{mmnp} for a detailed discussion). 

We now use the above trans--series formalism to compute the amplitudes $F^{(\ell)}$ for a cubic matrix model with potential
\be
V(z)=-z+{z^3\over 3}.
\ee
\noindent
A similar computation has been performed in \cite{mmnp} for the quartic matrix model. The recursion relation for the coefficients $r_n$ reads \cite{dfgz}
\be\label{oprecursion}
r_n\left(\sqrt{1-r_n-r_{n+1}}+\sqrt{1-r_n-r_{n-1}}\right)=g_s n.
\ee
\noindent
The continuum limit of equation \eqref{oprecursion} above is
\be\label{Rdifference}
R(t,g_s)\left(\sqrt{1-R(t,g_s)-R(t+g_s,g_s)}+\sqrt{1-R(t,g_s)-R(t-g_s,g_s)}\right)=t.
\ee
\noindent
At lowest order in $\ell$ and $g_s$ we find
\be
2R^{(0)}_0(t)\sqrt{1-2R^{(0)}_0(t)}=t.
\ee
\noindent
It turns out to be convenient to express everything in terms of a new variable, $R_{0}^{(0)}(t)\equiv r$. Power series expanding equation \eqref{Rdifference}, and solving it recursively, one can compute in this way the coefficients $R_{n}^{(\ell)}(r(t))$. The first few read
\be
\ba
R^{(0)}_2(t) &= -\frac{ (9 r-5)}{32 (1-3 r)^4}, \qquad &R^{(0)}_4(t) &= -\frac{3  \left(162 r^3+1017 r^2-1316 r+385\right)}{2048 (3 r-1)^9},\\
R^{(1)}_0(t) &= \frac{\sqrt{r}}{\sqrt[4]{3 r-1}}, \qquad &R^{(1)}_1(t) &= \frac{9 r^2-12 r+8}{192 (1-3r)^{5\over 2}}.
\ea
\ee

With these functions in hand, we can now compute the free energies using \eqref{FlogR}, up to an overall constant $C^{\ell}$ for each $F^{(\ell)}$. In turn, this can be compared to our prediction \eqref{multiformula}. We find
\be\label{multicubicthree}
\ba
F^{(1)} &= -{\sqrt{r}\over 4(3r-1)^{5\over 4}}\, \biggl\{ 1 + g_s {-8-228r+423r^2 \over 192 r(1-3r)^{5\over 2}} + \CO(g_s^2) \biggr\},\\
F^{(2)} &= -{r\over 32(3r-1)^{5\over 2}}\, \biggl\{ 1 + g_s {-8-228r+429r^2 \over 96 r(1-3r)^{5\over 2}} + \CO(g_s^2) \biggr\},\\
F^{(3)} &= -{r^{3\over 2}\over 192 (3r-1)^{15\over 4}}\, \biggl\{ 1 + g_s {-8-228r+429r^2\over 64 r(1-3r)^{5\over 2}} + \CO(g_s^2)\biggr\}.
\ea
\ee
\noindent
These results precisely have the structure predicted by (\ref{multiformula}). This can be explicitly checked by computing the quantities $\p_s\widehat{F}_1$, $\p^3_s\widehat{F}_0$ and $\hat{q}=\re^{\p_s^2\widehat{F}_0}$ from our analytical expressions \eqref{result}, \eqref{zerocubicder} and \eqref{onecubicder}, specialized to the cubic model. We find
\be\label{cubicquants}
\ba
\hat{q} &= -{r\over32\left(1-3r\right)^{5\over 2}}, \\
{1\over 6} \p^3_s\widehat{F}_0 &= \frac{-183 r^2+84 r+8}{96 r(1-3 r)^{5\over 2}},\\
\p_s \widehat{F}_1 &= \frac{-57 r^2+60 r-8}{192 r(1-3 r)^{5\over 2}}.
\ea
\ee
\noindent
If we now plug these in \eqref{multiformula} we immediately recover (\ref{multicubicthree}), up to an $r$--independent normalization of $C^{\ell}$. This is a rather strong check of our multi--instanton formulae.
 
\subsection{Multi--Instantons in 2d Quantum Gravity}

Unfortunately, the results we have just obtained for the cubic matrix model test the structure (\ref{multiformula}) only for low instanton number $\ell$. On the other hand, it is well--known (see, \textit{e.g.}, \cite{dfgz}) that the cubic matrix model has a double--scaling limit which describes pure 2d gravity. We shall now derive analytic expressions for $F^{(\ell)}$ in this limit, up to two loops, and for all $\ell \ge1$, by using the string equation, and we will verify that they are in perfect agreement with (\ref{multiformula}). 

The critical point of the cubic matrix model occurs at
\be
z_c={2\over 3\sqrt{3}}, \qquad r_c={1\over 3}. 
\ee
\noindent
The double--scaling limit is defined as
\be
g_s \rightarrow 0, \qquad z \rightarrow z_c, \qquad \kappa^{5\over 4}=(z_c-z)^{5\over 4}3^{5\over 8}g_s^{-1} \quad \mathrm{fixed}.
\ee
\noindent
In this limit, the specific heat of 2d gravity
\be
u(\kappa)=\left(3R(\kappa)-1\right)g_s^{-{2\over 5}}  
\ee
\noindent
satisfies the Painlev\'e I equation
\be
u(\kappa)^2-{1\over 6}u''(\kappa)=\kappa,
\ee
\noindent
as can be deduced from the difference equation \eqref{Rdifference}. The multi--instanton contributions can thus be obtained by finding a trans--series solution 
\be\label{painone}
u(z)=\sum_{\ell \ge 0} \zeta^{\ell} u^{(\ell)} (z)
\ee
\noindent
to the Painlev\'e I equation. One obtains a set of recursion equations for the $u^{(\ell)}(z)$ of the form
\be\label{recur}
-{1\over 6} (u^{(\ell)})'' + \sum_{k=0}^\ell u^{(k)} u^{(\ell-k)}=0, \qquad \ell \ge 1.
\ee
\noindent
Changing variables to
\be
z^{-{5\over 4}}={\sqrt{3}}x,
\ee
\noindent
each $u^{(\ell)}(x)$ has the structure 
\be\label{eansatz}
u^{(\ell)}(x) = u^{(\ell)}_0\, x^{a_{\ell}}\, \re^{-\ell A/x}\, \epsilon^{(\ell)}(x),
\ee
\noindent
where
\be
A = {8\over 5} \qquad \mathrm{and} \qquad \epsilon^{(\ell)}(x) = 1 + \sum_{k=1}^{\infty} u^{(\ell)}_k x^k.
\ee
\noindent
In particular, for $\ell=0,1$, we have
\be\label{firstwo}
\epsilon^{(0)}(x) = 1 - {x^2\over 16} + \CO(x^4), \qquad \epsilon^{(1)}(x) = 1 - {5 x\over 64} + \CO(x^2).
\ee

In the trans--series solution (\ref{painone}) $\zeta$ parameterizes the nonperturbative ambiguity. In principle the value of $u^{(1)}_0$ is not fixed by the recursion and it could be absorbed in $\zeta$. However, if we choose $\zeta$ to be identical to the nonperturbative ambiguity appearing in the matrix model, then the value of $u^{(1)}_0$ can be fixed, as we shall do in the following. 

From $\epsilon^{(\ell)}(x)$ it is possible to compute the multi--instanton corrections to the double--scaled free energy, $F^{(\ell)}_{\rm ds}(x)$, which are defined by
\be
u(z) =-F''_{\rm ds}(z), 
\ee
\noindent
and
\be
F_{\rm ds}(z) =\sum_{\ell=0}^{\infty} \zeta^{\ell}\, \re^{-\ell A /x} F_{\rm ds}^{(\ell)}(z). 
\ee
\noindent
As a function of $x$ they have the structure
\be\label{flst}
F^{(\ell)}_{\rm ds}(x) = f^{(\ell)}_0\, x^{\lambda_{\ell}}\, \Bigl\{ 1 +\sum_{k=1}^{\infty} f_k^{(\ell)} x^{\ell} \Bigr\}.
\ee
\noindent
The exponent $\lambda_\ell$ and coefficients $f^{(\ell)}_k$ can be related to the data appearing in $u^{(\ell)}$. We have, for example, 
\be\label{frels}
\lambda_{\ell}=a_\ell +{2\over 5}, \qquad f^{(\ell)}_0 = -{3^{1/5} u_0^{(\ell)} \over 12 \ell^2} \qquad \mathrm{and} \qquad f^{(\ell)}_1 =u^{(\ell)}_1 -{5\over 8 }  +{1 \over 8\ell}.
\ee

We can now use the equations (\ref{recur}) to obtain recursive relations for the various coefficients appearing in the ans\"atz (\ref{eansatz}). The recursion for $a_\ell$ is very simple,
\be
a_\ell ={2\over 5}+a_k + a_{\ell-k}, \qquad k =1, \ldots, \ell-1. 
\ee
\noindent
This requires $a_\ell$ to be linear in $\ell$, and we find 
\be\label{linearan}
a_\ell={\ell \over 2}-{2\over 5}, \qquad \lambda_{\ell}={\ell \over 2}.
\ee
%
%
%
\noindent
In order to find $F^{(\ell)}_{\rm ds}(x)$ up to two loops, we have to solve the recursion relations for both $u^{(\ell)}_0$ and $u^{(\ell)}_1$. Starting with $u^{(\ell)}_0$, we find the recursion
\be\label{u0rec}
u_0^{(\ell)} = {3^{1\over 5} \over 2(\ell^2-1)}\, \sum_{k=1}^{\ell-1} u_0^{(k)} u_0^{(\ell-k)}, \qquad \ell \ge 2.
\ee
\noindent
Using the well--known identity
\be
\sum_{k=1}^{\ell-1} k \left(\ell-k\right) = {1\over 6} \ell \left(\ell^2-1\right)
\ee
\noindent
one can immediately verify that
\be
\label{uoex}
u_0^{(\ell)} = {12\, \ell \over 4^{\ell} 3^{{4\ell +1 \over 5}}} \left( u_0^{(1)} \right)^\ell, \qquad \ell \ge 1,
\ee
\noindent
solves (\ref{u0rec}). The solution (\ref{uoex}) was already obtained in \cite{fy} in their study of nonperturbative effects in 2d gravity, and in \cite{cc} in their study of the 
trans--series solution to the Painlev\'e I equation. The second nontrivial coefficient in the trans--series solution, $u^{(\ell)}_1$, satisfies the recursion
\be\label{u1rec}
u^{(\ell)}_1 = {1\over 1-\ell^2}\, \Bigl( { 5 \ell (\ell-1) \over 8} - {12 \over \ell} \sum_{k=1}^{\ell-1} k (\ell-k)\, u^{(k)}_1 \Bigr), \quad \ell\ge 2.
\ee
\noindent
It can be easily checked that 
\be\label{u1sol}
u^{(\ell)}_1 = \Bigl( u^{(1)}_1-{47\over 96} \Bigr)\, \ell + {5\over 8} - {1\over 8 \ell}, \qquad \ell\ge 2,
\ee
\noindent
solves (\ref{u0rec}). Here one uses that 
\be
\sum_{k=1}^{\ell-1} k^2 \left(\ell-k\right) = {1\over 12} \ell^2 \left(\ell^2-1\right).
\ee
\noindent
Notice that (\ref{u1sol}) does not specialize to $u^{(1)}_1$ for $\ell=1$. Plugging these results in (\ref{frels}) we find
\be\label{dfval}
\ba
f^{(1)}_0 &=-{u^{(1)}_0 \over 4\cdot 3^{4\over 5}}, \\
f^{(\ell)}_0 &={(-1)^{\ell-1} \over \ell} \left( f^{(1)}_0 \right)^{\ell}, \qquad \ell \ge 1,\\
f^{(1)}_1 &=u^{(1)}_1-{1\over 2},\\
f^{(\ell)}_1 & =\Bigl( u^{(1)}_1-{47\over 96} \Bigr)\, \ell, \qquad \ell\ge 2.
\ea
\ee
\noindent
We can then obtain, directly from the string equation, 
\be\label{dsmultiformula}
\ba 
F^{(1)}_{\rm ds}(x) &= x^{1/2} f^{(1)}_0 \biggl\{ 1 - \Bigl({1\over 2}-u^{(1)}_1\Bigr)\, x + \CO(x^2) \biggr\}, \\
F^{(\ell)}_{\rm ds}(x) &= {(-1)^{\ell-1}\, x^{\ell/2} \over \ell} \left( f^{(1)}_0 \right)^{\ell}\, \biggl\{ 1 - \ell\, \Bigl( {47\over 96}-u^{(1)}_1 \Bigr)\, x +\CO(x^2) \biggr\}, \qquad \ell\ge 2.
\ea
\ee
\noindent
On the other hand, the double--scaling limit of our multi--instanton formulae (\ref{multiformula}) leads to
\be\label{dsexmultiformula}
\ba 
F^{(1)}_{\rm ds}(x) &= x^{1/2}\, \hat q_{\rm ds}^{1/2}\, \biggl\{ 1 - \Bigl( \phi_{1,1} + {1 \over 6} \phi_{0,3} \Bigr)\, x + \CO(x^2) \biggr\}, \\
F^{(\ell)}_{\rm ds}(x) &= {(-1)^{\ell-1}\, x^{\ell/2} \over \ell}\, \hat q_{\rm ds}^{\ell/2}\, \biggl\{ 1 - \ell\, \Bigl( \phi_{1,1} + {1 \over 6} \phi_{0,3} + \hat q_{\rm ds} \Bigr)\, x + \CO(x^2) \biggr\}, \qquad \ell\ge 2.
\ea
\ee
\noindent
In these formulae, $\hat q_{\rm ds}$, $\phi_{0,3}$ and $\phi_{1,1}$ are the double--scaling limits of $\hat q$, $\partial_s^3 \widehat F_0$ and $\partial_s \widehat F_1$, respectively. Their values can be obtained explicitly from (\ref{cubicquants}) and read
\be
\hat q_{\rm ds}=-{1\over 96}, \qquad \phi_{0,3}={47\over 16} \qquad \mathrm{and} \qquad \phi_{1,1}={17\over 192}.
\ee
\noindent
Since $u^{(1)}_1=-\frac{5}{64}$ (see (\ref{firstwo})), we see that the explicit solution to the recursion equations, (\ref{dsmultiformula}), agrees with the prediction (\ref{dsexmultiformula}) arising from our multi--cut analysis. Agreement of both formulae further requires
\be
f^{(1)}_0 = \Bigl( {\hat q_{\rm ds}\over 2\pi} \Bigr)^{1/2}.
\ee
\noindent
This is the normalization condition that fixes the value of $u^{(1)}_0$ so that the nonperturbative ambiguities are the same, as we pointed out above.

It is easy to solve the string equation to higher orders and we can use the results obtained in this way to give further tests of our general formulae. Since the $\ell$--instanton amplitude  (\ref{prediction}) gives contributions to the free energy at $\ell (\ell-1)/2$ loops, we will focus on the nonperturbative partition function 
\be
Z^{\rm np} = 1+\sum_{\ell\ge 1} z^\ell\, Z^{(\ell)}_{\rm ds} = \exp \left[ \sum_{\ell=1}^{\infty} z^\ell\, F^{(\ell)}_{\rm ds} \right], 
\ee
\noindent
where
\be
z=\zeta\, \re^{-A/x}.
\ee
\noindent
The general formula (\ref{prediction}) leads, in the double--scaling limit, to 
\be\label{dsprediction}
Z^{(\ell)}_{\rm ds} = {x^{\ell^2/2} \over (2\pi)^{\ell\over 2}}\, G_2(\ell+1)\, \hat q_{\rm ds}^{\ell^2/2}\, \biggl\{ 1 - x\, \Bigl( \ell\, \phi_{1,1} + {\ell^3 \over 6}\, \phi_{0,3} \Bigr) + \CO(x^2) \biggr\}.
\ee

We shall now present a slightly different reorganization of $Z^{\rm np}$, inspired by the analysis of \cite{cc}. Instead of considering the partition function $Z^{(\ell)}_{\rm ds}$ at fixed instanton number and expanded in $x$, we shall consider in $Z^{\rm np}$ the terms with a fixed power of $x$ but for \textit{all} instanton numbers. In other words, we write
\be\label{reorg}
Z^{\rm np}=\sum_{k\ge 0} x^k\, Z_k (\xi),
\ee
\noindent
where
\be
\xi = f_0^{(1)}\, x^{1/2}\, \re^{-A/x}
\ee
\noindent
and
\be
Z_k (\xi)=\delta_{k0}+\sum_{\ell \ge 1} z_{k,\ell}\, \xi^\ell.
\ee
\noindent
We are now going to extract a structural result for $Z_k(\xi)$ from the general formula (\ref{dsprediction}). This formula, based on the connection to the multi--cut matrix model, says that for each $\ell$ the starting power of $x$ is $\ell^2/2$. But the total power of $x$ in the expression above, for each fixed $\ell$, is $k+\ell/2$. Therefore, (\ref{dsprediction}) says that, at fixed $k$, $z_{k,\ell}$ is different from zero only for the {\it finite} number of $\ell$'s such that
\be\label{ineq}
{\ell (\ell-1)\over 2}\le k. 
\ee
\noindent
Let us denote by $\ell_{\rm m}(k)$ the maximum number satisfying this bound. We have, for example,
\be
\ell_{\rm m}(0)=1, \qquad \ell_{\rm m}(1)=2, \qquad \ell_{\rm m}(2)=2, 
\ee
\noindent
and so on. We then conclude that
\be
Z_k (\xi)=\delta_{k0}+\sum_{\ell=1}^{\ell_{\rm m}(k)} z_{k,\ell}\, \xi^\ell
\ee
is a {\it polynomial} of degree $\ell_{\rm m}(k)$. We have computed $Z_k(\xi)$ from the recursion (\ref{recur}) at high $k$ and found that indeed they are polynomials in $\xi$ of the right degree. We list them up to $k=6$:
\be
\ba
Z_0(\xi) &= 1+\xi, \\
Z_1(\xi) &= -\frac{\xi \left(2 \xi +111\right)}{192}, \\
Z_2 (\xi) &= \frac{\xi \left(1048 \xi +19299\right)}{24576}, \\
Z_3(\xi) &= -\frac{\xi \left(160 \xi^2+11705160 \xi +114670521\right)}{70778880}, \\
Z_4(\xi) &= \frac{\xi \left(552320 \xi^2+12466492352 \xi +79686828333\right)}{18119393280}, \\
Z_5(\xi) &= -\frac{ \xi \left(331932480 \xi^2+3646348240864 \xi +17179325749341\right)}{1159641169920}, \\
Z_6(\xi) &= \frac{\xi \left(102400 \xi^3+15883648256640 \xi^2+105786318127916160 \xi +395885797054644519\right)}{6679533138739200}.
\ea
\ee
\noindent
The results for $Z_k(\xi)$ are in full agreement with the expected structure (\ref{dsprediction}). For example, if $k$ is such that (\ref{ineq}) is saturated for $\ell_m(k)$, it follows from (\ref{dsprediction}) that the highest power of $\xi$ in $Z_k(\xi)$ should be given by 
\be
\hat q_{\rm ds}^k\, G_2(\ell_m(k)+1). 
\ee
\noindent
This is indeed the case in all the examples we have tested, and it is furthermore a nontrivial test of our regularization (\ref{interzn}). It says that the string equation for pure 2d gravity ``knows'' about the Barnes function, and therefore about the Gaussian matrix model partition function which we have used to regularize the full partition function.

There are two interesting applications of our above results for the multi--instanton amplitudes in 2d gravity. First, the formulae (\ref{dsexmultiformula}) for $F^{(\ell)}_{\rm ds}$ should yield the partition function of $\ell$ ZZ--branes in 2d gravity (\textit{i.e.}, in the $(2,3)$ minimal model coupled to gravity) after including their \textit{full} back--reaction on the target geometry. As explained in \cite{annulusZZ}, these amplitudes are divergent in the continuum theory, but one should be able to regularize them by taking into account the fact that the $\ell$ ZZ--branes smooth out the pinched cycle of the Riemann surface describing the model \cite{ss}. In section 4.1 we performed the computation of the amplitudes starting from the theory in which the pinched cycle is completely smoothed out, \textit{i.e.}, the two--cut solution. When we expanded naively around the background with a pinched cycle, we found that the expansion was singular. However, we resolved the singularity in the matrix model computation by performing an exact Gaussian integration of the $\ell$ eigenvalues. This led to (\ref{interzn}) and eventually to (\ref{dsexmultiformula}), which, as we have seen, agrees with the explicit trans--series solution of Painlev\'e I. In the language of \cite{annulusZZ} this means that our matrix model computation appropriately incorporates the back--reaction of the ZZ--branes, and it is in contrast with the computation of \cite{st} where the interactions between identical ZZ--branes have not been taken into account.

The second application of our result is more mathematical. The reorganization of the instanton expansion in (\ref{reorg}) can, of course, also be made at the level of the free energy and the specific heat. Indeed, if we write
\be
u=x^{-2/5} \sum_{k=0}^{\infty} x^k\, u_k(\xi),
\ee
\noindent
it can be seen by plugging this ans\"atz in Painlev\'e I that the $u_k(\xi)$ satisfy differential equations and are given by {\it rational} functions \cite{cc}. It then follows that in the expansion of the free energy
\be
F_{\rm ds}= \sum_{k=0}^{\infty} x^k\, f_k(\xi)
\ee
\noindent
the $f_k(\xi)$ are also rational. But, after exponentiation, the partition function (\ref{reorg}) is written in terms of {\it polynomials} $Z_k(\xi)$, as we have argued. This is rather surprising from the point of view of the original Painlev\'e I equation and seems to be a new structural result for the trans--series solution to Painlev\'e I.

\sectiono{Conclusions and Outlook}

In this paper we have studied multi--instanton configurations in multi--cut matrix models. 
As we have seen, producing general, formal expressions for their amplitudes is rather straightforward, although on a general background there are subtleties related to the existence of tachyonic directions. The focus of our paper has thus been the analysis of such expressions in the one--cut limit, generalizing the one--instanton formulae from our earlier work \cite{msw}. This is also a limit in which the generic expression needs a regularization, and we have proposed a natural one which fits all available data on multi--instantons, in particular the multi--instanton amplitudes in 2d gravity.

There are various avenues for future research. Our study has focused for simplicity on two--cut models and their one--cut limit, since this has made possible detailed tests of our approach. General formulae for more than two cuts can similarly be obtained in a rather straightforward way (closely related formulae in the general case have already been written down in \cite{eynard}). It is also straightforward to generalize the results of section 4 and to consider multi--instantons in a one--cut model which is obtained as the limit of a multi--cut model with more than two cuts. In this way one could incorporate different types of instantons, corresponding to tunneling from the filled cut to the several different empty cuts. Another venue of research deals with extending our formulae for the $(2,3)$ minimal model to general minimal string theories. This would produce amplitudes for the interaction of an arbitrary number of ZZ--branes, which fully consider their back--reaction in the target geometry, greatly enhancing our understanding of multiple brane interactions, albeit in the noncritical context. Naturally, this also raises the (much harder) question of how to derive our results directly in the continuum Liouville theory.

What remains one of the most interesting open issues for us would be to understand and test the connection between the multi--instanton formulae in the multi--cut case and the large--order behavior of the perturbative amplitudes for a fixed multi--cut background, following our general analysis in section 2. A significant step towards efficiently computing the topological $1/N$ expansion of multi--cut models was very recently given in \cite{hkr08, alm08}, building on \cite{hk,gkmw}, and opening way to a large--order analysis, 
and this is a subject we hope to return in the future. This would generalize the one--cut study performed in \cite{msw} and would also have direct applications in topological string theory on toric Calabi--Yau manifolds.

\section*{Acknowledgments}
We would like to thank Vincent Bouchard, Bertrand Eynard, Albrecht Klemm and Sara Pasquetti for useful discussions and/or correspondence.



\appendix

\section{Elliptic Functions: Definitions, Integrals and Properties}

We follow the conventions in \cite{grbook}. The complete elliptic integral of the first kind is defined as
\be
K (m) = \int_0^1 \frac{\rmd t}{\sqrt{(1-t^2)(1-m\, t^2)}}.
\ee
\noindent
In here $m=k^2$ is the parameter of the complete elliptic integral and $k$ is the elliptic modulus. One also defines the complementary modulus as $k'^2=1-k^2$. The complete elliptic integral of the second kind is defined as
\be
E (m) = \int_0^1 \rmd t\, \sqrt{\frac{1-m\, t^2}{1-t^2}}.
\ee
\noindent
Both elliptic integrals, $K (m)$ and $E (m)$, have a branch cut discontinuity in the complex $m$ plane running from $1$ to $\infty$. The complete elliptic integral of the third kind is defined as
\be
\Pi ( n | m ) = \int_0^1 \frac{\rmd t}{(1-n\, t^2) \sqrt{(1-t^2)(1-m\, t^2)}}.
\ee
\noindent
In here we take $0<k^2<1$ and $n$ is the elliptic characteristic.

The complete elliptic integrals of first and second kind satisfy the Legendre relation
\be
E(k) K(k') + E(k') K(k) - K(k) K(k') = \frac{\pi}{2}.
\ee
\noindent
One sometimes writes $K(k')=K'(k)$ and $E(k')=E'(k)$, and denotes these functions as complementary elliptic integrals. The complete elliptic integrals of first and second kind also relate to each other via derivation, \textit{e.g.},
\bea
\frac{\rmd K}{\rmd k} &=& \frac{E(k) - k'^2 K(k)}{k k'^2}, \\
\frac{\rmd E}{\rmd k} &=& \frac{E(k) - K(k)}{k}.
\eea
\noindent
Useful properties satisfied by the complete elliptic integral of the third kind include
\bea
&&
\Pi (n|m) + \frac{n}{1-n}\, \frac{1-k^2}{k^2-n}\, \Pi \left. \left( \frac{k^2-n}{1-n} \right| m \right) = \frac{k^2}{k^2-n}\, K(m), \\
&&
\Pi (n|m) + \Pi \left. \left( \frac{k^2}{n} \right| m \right) = K(m) + \frac{\pi}{2} \sqrt{\frac{n}{(1-n)(n-k^2)}}.
\eea

Let us next consider the elliptic geometry given by the curve $y^2(z) = M^2(z)\, \sigma(z)$, with
\be
\sigma (z) \equiv \prod_{i=1}^4 \left( z-x_i \right) = (z-x_1) (z-x_2) (z-x_3) (z-x_4)
\ee
\noindent
and $x_1<x_2<x_3<x_4$. A non--constant single--valued function $f(z)$, with $z \in \BC$, is called an elliptic function if it has two periods $2\omega_1$ and $2\omega_2$, \textit{i.e.},
\be
f(z+2m\omega_1+2n\omega_2) = f(z),
\ee
\noindent
with $m,n \in \BZ$. The half--period ratio of the elliptic geometry, $\tau$, is defined as $\tau = \frac{\omega_1}{\omega_2}$, and the elliptic nome is defined as $q = \rme^{\rmi \pi \tau}$. The elliptic nome can be related to the complete elliptic integral of the first kind as
\be
q = \exp \left( - \pi \frac{K(k')}{K(k)} \right),
\ee
\noindent
making it clear that $q(m)$ has a branch cut discontinuity in the complex $m$ plane running from $1$ to $\infty$. The inverse relation expresses the complete elliptic integral of the first kind in terms of a Jacobi theta function as (here $\vartheta_3 (q) = \vartheta_3 (q|z=0)$)
\be
K(k) = \frac{\pi}{2}\, \vartheta_3^2 (q).
\ee

Useful integrals which are used in the main text, and which can be obtained from the tables in \cite{grbook}, are
\bea
\int_{x_{1}}^{x_{2}} \frac{\rmd z}{\sqrt{\sigma(z)}} &=& - \frac{2\rmi}{\sqrt{(x_4-x_2)(x_3-x_1)}}\, K(k), \\
\int_{x_{3}}^{x_{4}} \frac{\rmd z}{\sqrt{\sigma(z)}} &=& - \frac{2\rmi}{\sqrt{(x_4-x_2)(x_3-x_1)}}\, K(k),
\eea
\noindent
along the $A$--cycles; as well as
\be
\int_{x_{2}}^{x_{3}} \frac{\rmd z}{\sqrt{\sigma(z)}} = \frac{2}{\sqrt{(x_4-x_2)(x_3-x_1)}}\, K(k'),
\ee
\noindent
along the $B$--cycle. In the formulae above the elliptic modulus is
\be
k^2 = \frac{(x_1-x_2)(x_3-x_4)}{(x_1-x_3)(x_2-x_4)}.
\ee

\section{Expression for $\partial_s F_1$ in the Two--Cut Matrix Model}

A final ingredient required in the main body of the paper in order to explicitly write all multi--instanton formulae in the two--cut matrix model, in particular in order to compute the two--loops coefficient, is the derivative of genus one free--energy, $F_1(t_1,t_2)$. This was computed in \cite{kmt}, with the result
\be
F_1 = - \frac{1}{24} \sum_{i=1}^4 \ln M_i - \frac{1}{2} \ln K(k) - \frac{1}{12} \sum_{i<j} \ln \left( x_i-x_j \right)^2 + \frac{1}{8} \ln \left( x_1-x_3 \right)^2 + \frac{1}{8} \ln \left( x_2-x_4 \right)^2,
\ee
\noindent
where we now need to evaluate
\bea
- \frac{\partial F_1}{\partial s} &=& \frac{1}{24} \sum_{i=1}^4 \frac{1}{M_i} \frac{\partial M_i}{\partial s} + \frac{E(k) - k'^2 K(k)}{4 k^2 k'^2\, K(k)}\, \frac{\partial k^2}{\partial s} + \frac{1}{6} \sum_{i<j} \frac{1}{x_i-x_j} \left( \frac{\partial x_i}{\partial s} - \frac{\partial x_j}{\partial s} \right) - \nonumber \\
&& - \frac{1}{4 \left( x_1-x_3 \right)} \left( \frac{\partial x_1}{\partial s} - \frac{\partial x_3}{\partial s} \right) - \frac{1}{4 \left( x_2-x_4 \right)} \left( \frac{\partial x_2}{\partial s} - \frac{\partial x_4}{\partial s} \right).
\eea
\noindent
The only novelty is $\frac{\partial M_i}{\partial s}$, which may be computed with some of the formulae we presented in the main body of the paper. The result is
\bea
\frac{1}{24} \sum_{i=1}^4 \frac{1}{M_i} \frac{\partial M_i}{\partial s} &=& \frac{\pi \sqrt{(x_1-x_3) (x_2-x_4)}}{8\, K(k)}\, \frac{1}{\prod_{i<j} \left(x_i-x_j \right)} \left( \frac{M_1''}{M_1^2}\, (x_2-x_3)(x_2-x_4)(x_3-x_4) - \right. \nonumber \\
&& - \frac{M_2''}{M_2^2}\, (x_1-x_3)(x_1-x_4)(x_3-x_4) + \frac{M_3''}{M_3^2}\, (x_1-x_2)(x_1-x_4)(x_2-x_4) - \nonumber \\
&& \left. - \frac{M_4''}{M_4^2}\, (x_1-x_2)(x_1-x_3)(x_2-x_3) \right).
\eea
\noindent
Assembling all different pieces together, it finally follows
\bea\label{dsf12cut}
- \frac{\partial F_1}{\partial s} &=& \frac{\pi}{8 K(k)}\, \frac{\sqrt{(x_1-x_3) (x_2-x_4)}}{\prod_{i<j} \left(x_i-x_j \right)} \left( \frac{M_1''}{M_1^2}\, (x_2-x_3)(x_2-x_4)(x_3-x_4) - \right. \nonumber \\
&& - \frac{M_2''}{M_2^2}\, (x_1-x_3)(x_1-x_4)(x_3-x_4) + \frac{M_3''}{M_3^2}\, (x_1-x_2)(x_1-x_4)(x_2-x_4) - \nonumber \\
&& \left. - \frac{M_4''}{M_4^2}\, (x_1-x_2)(x_1-x_3)(x_2-x_3) \right) + \frac{\pi \left( E(k) - k'^2 K(k) \right)}{2 K^2(k)}\, \frac{(x_1-x_3)^{3/2} (x_2-x_4)^{3/2}}{\prod_{i<j} \left(x_i-x_j \right)^2} \cdot \nonumber \\
&& \cdot \left( \frac{1}{M_1}\, (x_2-x_3)^2 (x_2-x_4)^2 (x_3-x_4)^2 + \frac{1}{M_2}\, (x_1-x_3)^2 (x_1-x_4)^2 (x_3-x_4)^2 + \right. \nonumber \\
&& \left. + \frac{1}{M_3}\, (x_1-x_2)^2 (x_1-x_4)^2 (x_2-x_4)^2 + \frac{1}{M_4}\, (x_1-x_2)^2 (x_1-x_3)^2 (x_2-x_3)^2 \right) + \nonumber \\
&& + \frac{\pi \sqrt{(x_1-x_3) (x_2-x_4)}}{6 K(k)} \left( \frac{3 x_1^2 - x_1 x_2 - x_1 x_4 - x_2 x_4 - 4 x_1 x_3 + 2 x_2 x_3 + 2 x_3 x_4}{(x_1-x_2)^2 (x_1-x_3)^2 (x_1-x_4)^2 M_1} + \right. \nonumber \\
&& + \frac{3 x_2^2 - x_1 x_2 - x_1 x_3 - x_2 x_3 - 4 x_2 x_4 + 2 x_1 x_4 + 2 x_3 x_4}{(x_1-x_2)^2 (x_2-x_3)^2 (x_2-x_4)^2 M_2} + \nonumber \\
&& + \frac{3 x_3^2 - x_2 x_3 - x_2 x_4 - x_3 x_4 - 4 x_1 x_3 + 2 x_1 x_2 + 2 x_1 x_4}{(x_1-x_3)^2 (x_2-x_3)^2 (x_3-x_4)^2 M_3} + \nonumber \\
&& \left. + \frac{3 x_4^2 - x_1 x_3 - x_1 x_4 - x_3 x_4 - 4 x_2 x_4 + 2 x_1 x_2 + 2 x_2 x_3}{(x_1-x_4)^2 (x_2-x_4)^2 (x_3-x_4)^2 M_4} \right).
\eea

\section{Series Expansion of the Vanishing Period $t_2$}

In the two--cut matrix model, the vanishing period, \textit{i.e.}, the period of the differential $y(x)\, \rd x$ which vanishes as $x_4 \rightarrow x_3$, is given by the following integral:
\be 
t_2 = {1\over 2\pi} \int_{x_3}^{x_4} \rd x\, M(x) {\sqrt{|\sigma(x)|}}. 
\ee
\noindent
We want to find a systematic series expansion of this quantity in powers of 
\be
\epsilon=x_4-x_3. 
\ee
\noindent
This is easily done by writing
\be
(x-x_1)(x-x_2) = (x_3-x_1)(x_4-x_2) \Bigl( 1 + {x-x_3 \over x_3-x_1} \Bigr) \Bigl( 1 + {x-x_4 \over x_4 -x_2} \Bigr). 
\ee
\noindent
If we now introduce the variable
\be
u={x-x_3 \over \epsilon}
\ee
\noindent
the integral defining $t_2$ becomes
\be
\ba
t_2 = & {\epsilon^2 \over 2\pi}\, M(x_3) \left[ (x_3-x_1) (x_3-x_2+\epsilon) \right]^{1\over 2} \cdot \\ 
& \qquad \cdot \int_{0}^{\epsilon} \rd u\, {M(x_3 + \epsilon u) \over M(x_3)}\, \Bigl( 1 + \epsilon {u \over x_3-x_1} \Bigr)^{1\over 2} \Bigl( 1 + \epsilon {u - 1 \over \epsilon + x_3 -x_2} \Bigr)^{1\over 2} {\sqrt{u(1-u)}}.
\ea
\ee
\noindent
This expression admits a systematic expansion in $\epsilon$. In order to go to two loops in the computation of multi--instanton effects, we need to compute $t_2$ to order $\epsilon^4$. One finds, after setting $x_3 \equiv x_0$,
\be\label{epstwo}
\ba
t_2 &= {\epsilon^2 \over 16}\, M(x_0) \sqrt{(x_0-x_1)(x_0-x_2)}\, \biggl[  1 + {\epsilon \over 4} \biggl( {1 \over x_0 - x_1} + {1 \over x_0 - x_2} + {2 M'(x_0) \over M(x_0)} \biggr) - \\
& - {5 \epsilon^2 \over 128}\, \biggl( {(x_1-x_2)^2 \over (x_0-x_1)^2 (x_0-x_2)^2} + {4  (x_1+x_2-2 x_0) M'(x_0) \over (x_0-x_1)(x_0-x_2) M(x_0)} - 4 {M''(x_0) \over M(x_0)} \biggr) + \cdots \biggr].
\ea
\ee

\section{Relating Moment Functions of One and Two--Cut Models}

In this appendix we wish to relate two--cut moments to one--cut moments, following the line which we outlined in the main body of the paper. In the one--cut limit, one makes use of $M_{1} (z) = M(z) (z-x_0)$ and may write, for $z \not = x_0$
\bea
M(z) &=& - \frac{M_{1}(z)}{x_0-z}, \\
M'(z) &=& - \frac{1}{x_0-z} \left( M_{1}'(z) + \frac{M_{1}(z)}{x_0-z} \right), \\
M''(z) &=& - \frac{1}{x_0-z} \left( M_{1}''(z) + \frac{2}{x_0-z} \left( M_{1}'(z) + \frac{M_{1}(z)}{x_0-z} \right) \right),
\eea
\noindent
alongside with $M^{(n)}(x_0) = \frac{1}{n+1}\, M_1^{(n+1)}(x_0)$, $n\ge 0$, among several other similar expressions.

\vfill

\eject


\bibliographystyle{plain}

\end{document}